\documentclass{aa}  
\usepackage[switch]{lineno} 

\usepackage{graphicx}
\usepackage{txfonts}
\usepackage{xcolor}
\usepackage{ulem}
\usepackage{hyperref}
\usepackage{orcidlink}
\usepackage{float}
\usepackage{manyfoot}
\hypersetup{
  colorlinks,
  citecolor=blue,
  linkcolor=blue,
  urlcolor=blue}
\begin{document} 

   \title{Evolution of binaries containing a hot subdwarf and a white dwarf to double white dwarfs, and  double detonation supernovae with hypervelocity runaway stars}
   \titlerunning{Evolution of binaries containing a hot subdwarf and a white dwarf}

   \author{Abinaya Swaruba Rajamuthukumar\inst{1}\orcidlink{0000-0002-1872-0124}
        \and
        Evan B. Bauer\inst{2}\orcidlink{0000-0002-4791-6724}
        \and
        Stephen Justham\inst{1}\orcidlink{0000-0001-7969-1569}
        \and
        R\"udiger Pakmor\inst{1}\orcidlink{0000-0003-3308-2420}
        \and
        \\ Selma E. de Mink\inst{1}\orcidlink{0000-0001-9336-2825}
        \and
        Patrick Neunteufel\inst{1}\orcidlink{0000-0001-5853-6017}
        }
\institute{Max Planck Institut für Astrophysik, Karl-Schwarzschild-Straße 1, 85748 Garching bei München, Germany\\
              \email{abinaya@mpa-garching.mpg.de}
        \and
            Center for Astrophysics | Harvard \& Smithsonian, 60 Garden Street, Cambridge, MA 02138, USA
            }
    \authorrunning{A.S.Rajamuthukumar et al}

 
  \abstract
 {Compact binaries containing hot subdwarfs and white dwarfs have the potential to evolve into a variety of explosive transients. These systems could also explain hypervelocity runaway stars such as US 708. We use the detailed binary evolution code MESA to evolve hot subdwarf and white dwarf stars interacting in binaries. We explore their evolution toward double detonation supernovae, helium novae, or double white dwarfs. We present a grid of 3120 binary evolution models that map from initial conditions, such as the orbital period and masses of the hot subdwarf and white dwarf, to these outcomes. The minimum amount of helium required to ignite the helium shell that leads to a double detonation supernova in our grid is $\approx 0.05 \, \mathrm{M_{\odot}}$, likely too large to produce spectra similar to normal Type Ia supernovae, but compatible with inferred helium shell masses from some observed peculiar Type I supernovae. We also provide the helium shell masses for our double white dwarf systems, with a maximum He shell mass of $\approx 0.18\,\mathrm{M_{\odot}}$. In our double detonation systems, the orbital velocity of the surviving donor star ranges from  $\approx 450 \, \mathrm{km\,s^{-1}}$ to $\approx 1000 \, \mathrm{km\,s^{-1}}$.  Among the surviving donors, we also estimate the runaway velocities of proto-white dwarfs, which have higher runaway velocities than hot subdwarf stars of the same mass. Our grid will provide a first-order estimate of the potential outcomes for the observation of binaries containing hot subdwarfs and white dwarfs from future missions like Gaia, LSST, and LISA.}


   \maketitle
%

\section{Introduction} \label{sec:Introduction}

Hot subdwarfs are stripped helium (He) stars that burn He in their core with little to no hydrogen (H) envelope \citep{1986A&A...155...33H, 2016PASP..128h2001H}. Studies such as \cite{2002MNRAS.336..449H,2003MNRAS.341..669H} suggest that binary interactions such as stable or unstable mass transfer plays a major role in the formation of hot subdwarfs. Observations provide evidence that most of the hot subdwarfs are found with binary companions \citep[e.g.,][]{2020A&A...642A.180P}. Further, a significant fraction of hot subdwarfs are found in close orbits ($P_{\mathrm{orb}} \sim 10\, \text{days}$) with white dwarfs (WDs), which are likely post common envelope systems \citep{2002MNRAS.336..449H,2003MNRAS.341..669H}. \newline
\begin{figure*}
    \centering
    \includegraphics[trim = 0 0 0 0 ,width=1\textwidth]{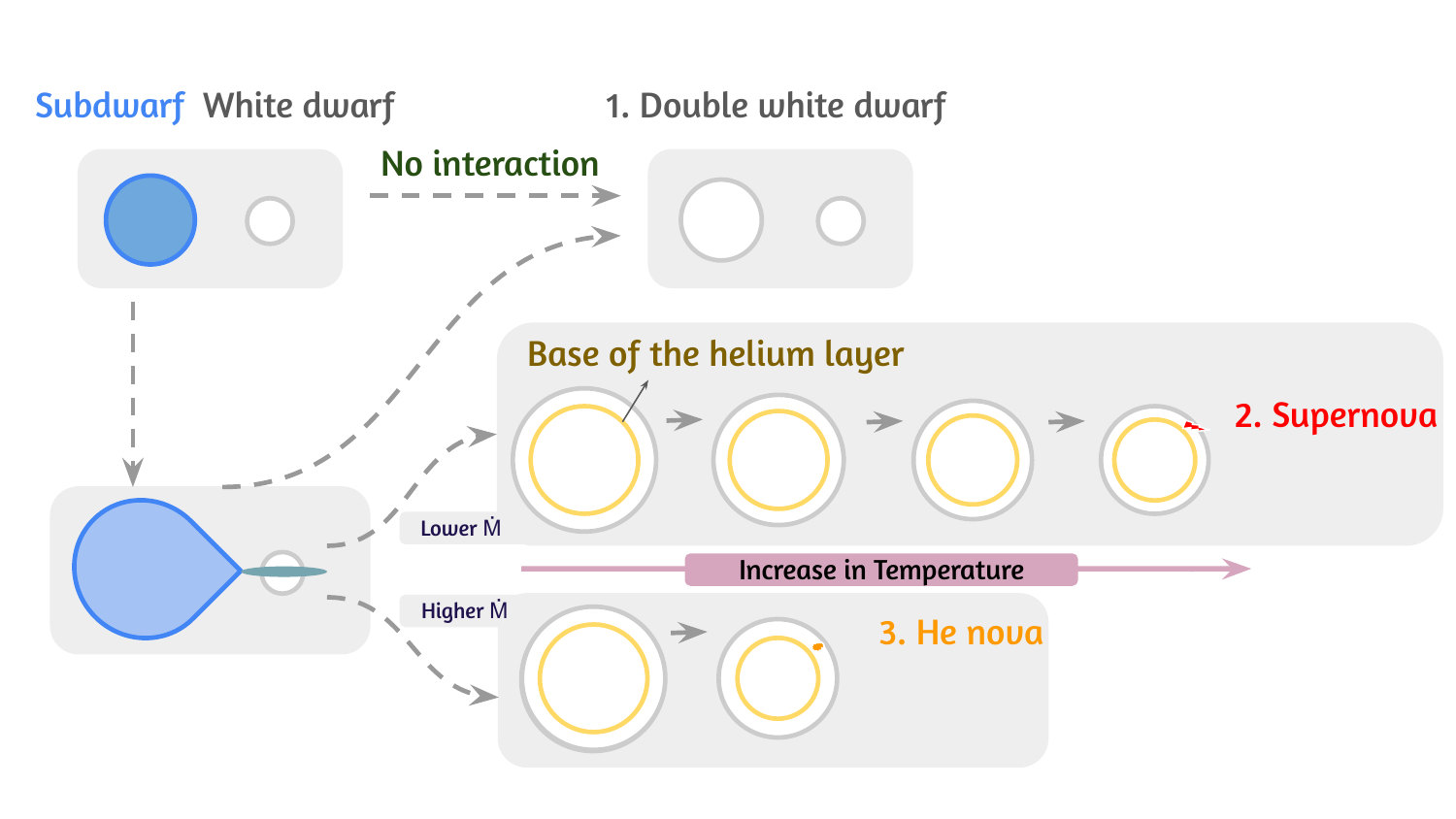}
    \caption{Schematic diagram representing possible evolution channels of a hot subdwarf + WD binary to a double white dwarf, a double detonation supernova or a He nova. Double detonation supernovae and He novae result from systems that undergo mass transfer and later experience thermonuclear instability in the He layer on the surface of the WD. The differentiation between double detonation supernova and He nova is based on the He layer's critical density (see Sect.~\ref{sec:Conditions for ignition}). Double white dwarfs result from systems that either evolve in isolation or involve episodes of accretion but do not enter unstable nuclear burning.} 
    \label{fig:cartoon}
\end{figure*}
Compact binaries containing hot subdwarfs and WDs have gained significant interest following the discovery of a super-Chandrasekhar binary by \citet{2000MNRAS.317L..41M}. This binary is particularly interesting as a potential Type Ia supernova progenitor (but, see also \citealt{2001A&A...376L...9E} for other outcomes).  
The number of discoveries of binaries containing hot subdwarfs with WD companions has steadily increased \citep[e.g.,][]{2004Ap&SS.291..321N,2012ApJ...759L..25V, Geier2013, 2023MNRAS.525..183S}. Large spectroscopic studies have been characterizing binaries containing hot subdwarfs \citep[e.g.,][]{2022A&A...666A.182S}. In addition to those detached binaries, \cite{2020ApJ...898L..25K, 2020ApJ...891...45K} identified binaries in which the hot subdwarf is Roche-lobe filling. Furthermore, some of recent the observed binaries such as CD-30°11223 \citep{2012ApJ...759L..25V,Geier2013} and HD 265435 \citep{2021NatAs...5.1052P} are verification binaries for LISA (for more details see \citealt{2024ApJ...963..100K}) and suggested progenitors of double donation supernovae \citep{2022ApJ...925L..12K}.

Interacting hot subdwarf + WD binaries could potentially result either in an explosion only in the He shell of the WD (a He nova,  \citealt{2000IAUC.7552....1K,2010ApJ...720..581G,2014MNRAS.445.3239P}) or a thermonuclear explosion of a WD (probably as a supernova Type Ia, \citealt{1973ApJ...186.1007W,1984ApJS...54..335I}). The thermonuclear explosion of the WD could result from a Chandrasekhar mass explosion \citep{2003A&A...412L..53Y,2009MNRAS.395..847W,2016ApJ...821...28B} or from a sub-Chandrasekhar mass double detonation explosion due to He accretion. In the double detonation mechanism, He-accreting WDs undergo the first detonation in the He shell that triggers the second detonation in the C/O core leading to a thermonuclear supernova.

The He donor channel as a mechanism to produce a double detonation was already investigated in the early 1980s by \cite{1980ApJ...237..142T,1980ApJ...242..749T}, \cite{1982ApJ...253..798N,1982ApJ...257..780N}.  Later studies by \cite{1990ApJ...354L..53L}, \cite{1990ApJ...361..244L}, \cite{1991A&A...252..669L}, \cite{1991ApJ...370..272L}, \cite{2007A&A...476.1133F}, and \cite{2010A&A...514A..53F} further explored He-accreting WDs and established that they are promising channels for SNe Ia. However, they also estimated that single degenerate double detonations require thick He shells which likely will lead to spectra representing a peculiar supernova Ia. The light curve of ZTF18aaqeasu (also known as SN 2018byg and ATLAS 18pqq), \cite{2019ApJ...873L..18D}, suggests that it is a double detonation supernova with a massive He shell ($\sim 0.15\,\mathrm{M_{\odot}}$). This thick He shell aligns with predictions by \cite{2017ApJ...845...97B} for a hot subdwarf + WD binary (CD-30°11223). On the contrary, studies by \cite{2010ApJ...714L..52S}, \cite{2010ApJ...719.1067K}, and \cite{2011ApJ...734...38W} also explored channels that would enable the synthetic spectra to resemble a ``normal'' Type Ia supernova. Furthermore, \cite{2013A&A...559A..94W} and \cite{2016A&A...589A..43N} probe the He star + WD binary channels as a potential single-degenerate double detonation mechanism. In addition, some studies explored the effects of WD rotation \citep{2004A&A...419..623Y} and magnetic fields \citep{2017A&A...602A..55N} on He ignition. 

He donor channels are also interesting as a mechanism to produce hypervelocity runway stars such as US 708. Should there be a full explosion of the WD, the donor can be ejected at very high velocities \citep{2009A&A...493.1081J,2009A&A...508L..27W}. Depending on the time of explosion of the accretor the donor star can either be a He star or a white dwarf. This mechanism could explain some of the observed population of hypervelocity runaway stars with velocities of $\lesssim 1000\,\mathrm{km\,s^{-1}}$\citep{2015Sci...347.1126G,2015ApJ...804...49B,2020A&A...641A..52N,2022A&A...663A..91N}. 

Given the increasing number of observed hot subdwarf and white dwarf binaries, interest in understanding the fate of these systems has been growing. Our galaxy contains at least $10^{3} - 10^{4}$ hot subdwarf + WD binaries that are likely to interact within the subdwarf lifetime \citep{2021ApJ...922..245B}. Despite detailed modeling studies of individual systems, the literature still lacks a broad study of binary spatial configurations that lead to double detonation supernovae. The purpose of this paper is to map the initial binary configurations of hot subdwarf and WD binaries to their fates: double detonation supernovae, He novae, and double white dwarfs.

In this study, we use the 1D stellar evolution code \textsc{mesa} to simulate a dense grid of hot subdwarf + WD binaries. Unlike some of the previous studies, we also model the accreting WD. We simultaneously evolve both the donor and accretor, including binary interactions, aiming to address two main objectives: 1) Identifying the region of initial parameter space that distinguishes the fate of the systems, and determining whether they evolve into double detonation supernovae, He novae, or double white dwarfs, and characterize the properties of any surviving system or object 2) Investigating the velocities of the runaway hot subdwarf in the event of double detonation supernovae, thereby providing constraints to observed runaway stars.

The paper is structured as follows: In Section \ref{sec:Setup} we explain our methods. Section \ref{sec:Examples} illustrates examples of three distinct outcomes under consideration: a double detonation supernova, He nova, or a double white dwarf. We present the fate of the binaries across our parameter space in Section \ref{sec:Ignition regions across initial parameter space}. In Section \ref{sec:Ignition properties} we give more details about the final state of the binaries, and in particular, runaway velocities in \ref{sec:Runaway velocities}. We discuss the results in Section \ref{sec:Discussion} and conclude in Section \ref{sec:Conclusion}.

\section{Methods} \label{sec:Setup}

\begin{figure}
    \centering
     \includegraphics[width=1\columnwidth]{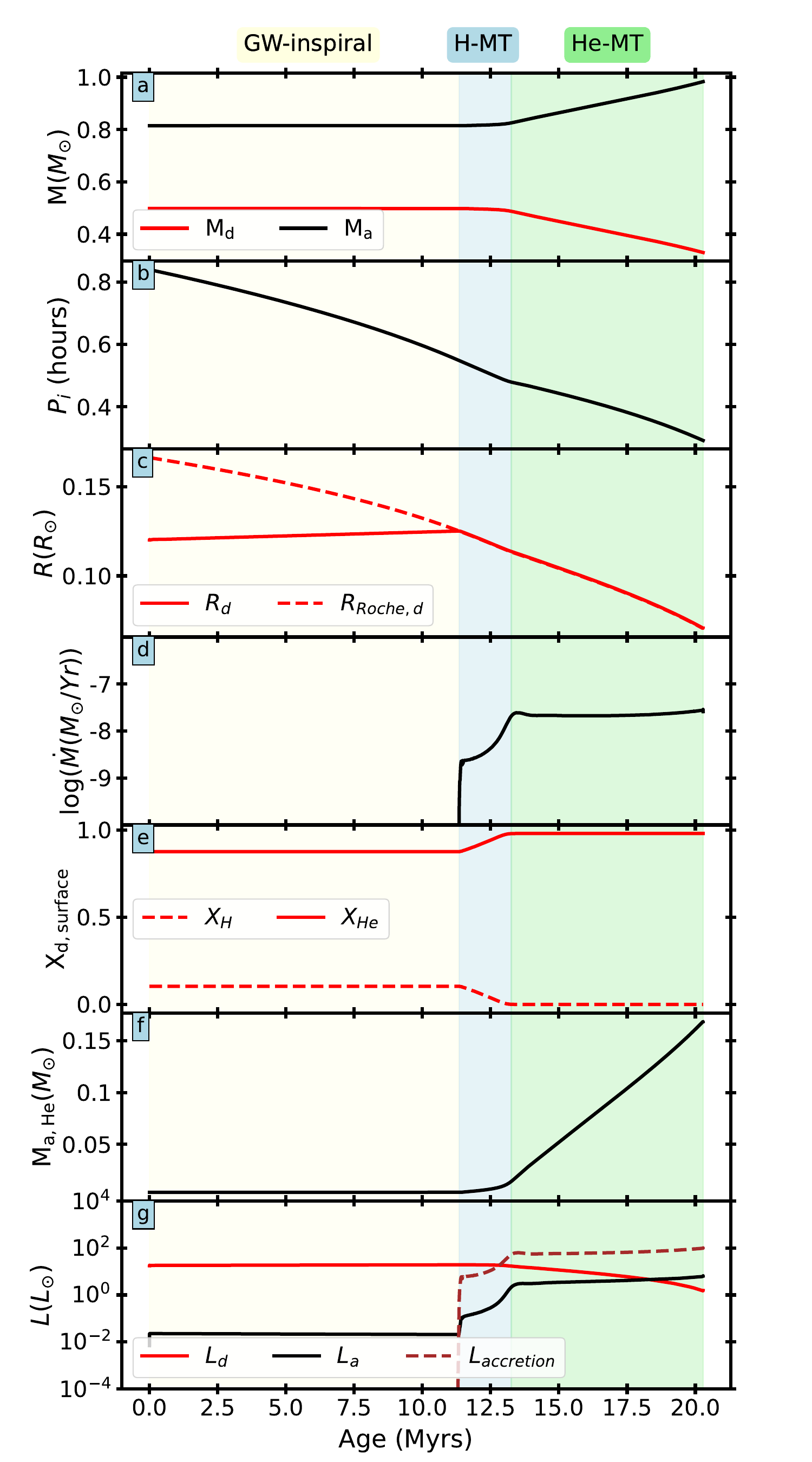}
    \caption{Example of a binary with $\mathrm{M_d = 0.5\,\mathrm{M_{\odot}}}$, $\mathrm{M_a = 0.8 \,\mathrm{M_{\odot}}}$, and $\mathrm{P_i = 0.84\, hours}$,  where a thermonuclear explosion occurs in the WD (accretor). The background colors yellow, blue, and green represent the gravitational-wave inspiral phase (GW-inspiral), H mass transfer phase (H-MT), and He mass transfer phase (He-MT), respectively.  Panels (a) through (f) show the evolution of various parameters: (a) Mass Evolution: $M_a$ (accretor) and $M_d$ (donor); (b) Orbital Period Evolution; (c) Radius Evolution: $R_d$ (donor's radius) and $R_{roche,d}$ (Roche radius of the donor); (d) mass transfer rate; (e) Evolution of surface mass fraction: $X_H$ (hydrogen) and $X_{He}$ (helium) in the donor; (f) He Mass on the Accretor; (g) Luminosity Evolution: $L_a$ (accretor), $L_d$ (donor), and $L_{\text{accretion}}$ (accretion). The accretor gains mass at the rate of $\sim 10^{-8} \, \mathrm{M_{\odot}yr^{-1}}$, resulting in the He ignition in the He layers denser than the assumed critical density for detonation ($> 10^{6} \, \mathrm{g\,cm^{-3}}$). }
    \label{fig:Explosion}
\end{figure}

\begin{figure*}
    \centering
    \includegraphics[width=\textwidth, trim=225 0 250 0, clip]{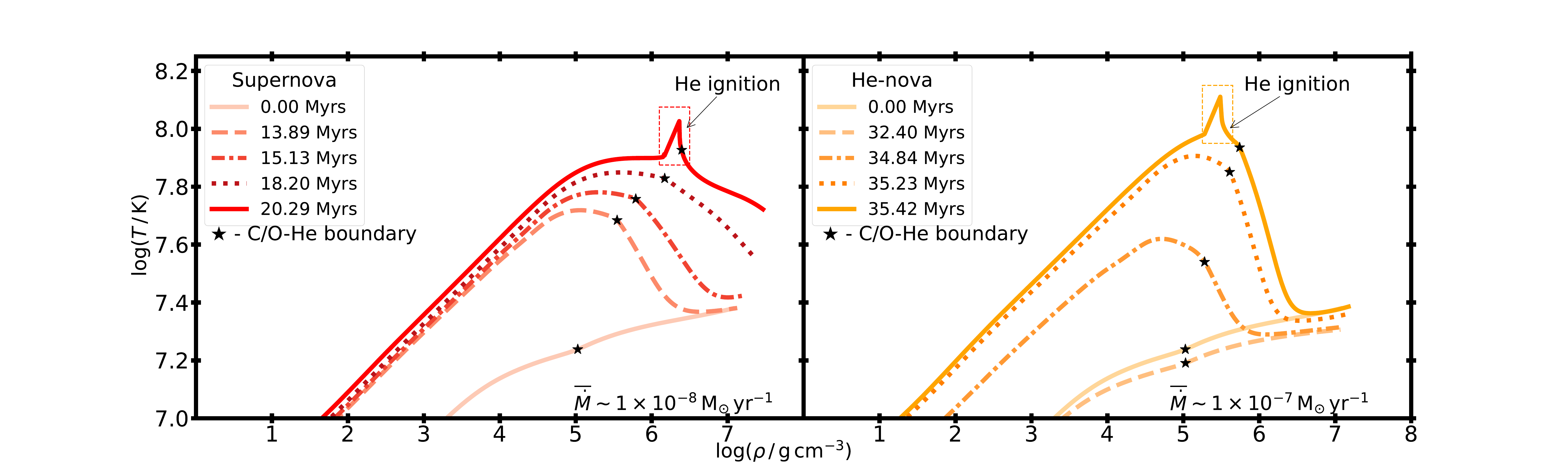}
    \caption{Each curve depicts the evolution of the temperature-density profile of the accreting white dwarf at various times, as indicated in the legend, for a system that undergoes double detonation supernova (left panel) and a He nova (right panel). The black star indicates the C/O-He boundary, that moves to higher density as the WD accretes and contracts. In the left panel, the WD accretes at the rate of $\sim 10^{-8} \, \mathrm{M_{\odot}yr^{-1}}$, igniting the high density regions of the He shell ($> 10^{6} \, \mathrm{g\,cm^{-3}}$), and so we classify the outcome as a double detonation supernova. In the right panel, the accretor's evolution is initially dominated by the cooling of the WD. After the accretion starts, the WD accretes at $\sim 10^{-7} \, \mathrm{M_{\odot}yr^{-1}}$. The WD experiences compressional heating, causing the low-density ($< 10^{6} \, \mathrm{g\,cm^{-3}}$) layers to ignite explosively. We then classify this system as a He nova.}
    \label{fig:TRho}
\end{figure*}

In this section, we describe the physical assumptions used in this study. We use Modules for Experiments in Stellar Astrophysics (modified\footnotemark[1]\footnotetext[1]
{\small The composition of the accretion stream is modified to change the mass fraction of H to He throughout accretion. See later in Sect.~\ref{sec:Setup} for more details. This is done by modifying the \texttt{binary\_mdot.f90} file. All files including the modified \texttt{binary\_mdot.f90} can be found at \href{https://zenodo.org/records/13473758}{Zenodo}.}
MESA version r23.05.1 \citealt{Paxton2011, Paxton2013, Paxton2015, 2018ApJS..234...34P, Paxton2019,2023ApJS..265...15J}) to model the binary evolution of hot subdwarf + WD binaries.

Convective Treatment: We use the \texttt{Schwarzschild} stability criterion to determine the regions that are unstable to convection. Further, mixing by convection is followed using the mixing length theory \citep{1958ZA.....46..108B}, with \texttt{mixing\_length\_alpha = 2} and \texttt{MLT\_option = 'TDC'}, the time-dependent convection model based on \cite{1986A&A...160..116K}. For convection treatments in the He core of hot subdwarf, we employ ``predictive mixing," which allows the core to grow while avoiding breathing pulses during the late stages of He core burning. For details on convection in hot subdwarfs in general, see \cite{2021MNRAS.503.4646O}.

Nuclear Network: When evolving the binary system, we use \texttt{basic\_plus\_fe56\_ni58.net} for hot subdwarf stars and a custom \texttt{nco.net} from \citealt{2017ApJ...845...97B} (included in the \href{https://zenodo.org/records/13473758}{Zenodo} repository) for WDs. For the hot subdwarf, the nuclear network \texttt{basic\_plus\_fe56\_ni58.net} contains stable isotopes $^{56}\text{Fe}$ and $^{58}\text{Ni}$ in addition to the basic isotopes $^{1}\text{H}, \, ^{3}\text{He}, \, ^{4}\text{He}, \, ^{12}\text{C}, \, ^{14}\text{N}, \, ^{16}\text{O}, \, ^{20}\text{Ne}, \, ^{24}\text{Mg}$. This has been chosen to account for all isotopes and reactions from He burning. For the WD, the nuclear network includes the NCO ($^{14}$N$(e^-,\nu)^{14}$C$(\alpha,\gamma)^{18}$O)  chain, in which the electron capture to $^{14}\text{N}$ in high densities ($\sim 10^6 \, \mathrm{g \, cm^{-3}}$), leads to the formation of $^{14}\text{C}$. This $^{14}\text{C}$ can further capture an alpha particle inducing a thermonuclear runaway in the accretor \citep{2017ApJ...845...97B}. 

Stellar winds: To account for mass loss during the red giant branch, we use the Reimers prescription \citep{1975MSRSL...8..369R} from MESA and switch to the  Blocker prescription \citep{1995A&A...299..755B} during the asymptotic giant branch phase. Further, we assume no winds during the hot subdwarf phase.  The mass and luminosity of these low-mass stripped stars are too low to drive strong winds that would affect the binary evolution and hence the results \citep{2016A&A...593A.101K}. 

Hot subdwarf: A hot subdwarf star is a low mass core He burning star with a very thin ($ \lesssim 10^{-2} \,\mathrm{M_{\odot}}$; \citealt{2016PASP..128h2001H}) H envelope.  To create starting models for hot subdwarfs, we first evolve single stars from the pre-main sequence until He ignition. The hot subdwarf gets stripped either during a mass transfer phase or a CE phase. In this work, we approximate the mass loss during the mass-transfer or CE phase by applying a high wind mass loss rate, while freezing the nuclear reactions in the core. The stripping continues until the total mass of H in the envelope is $3 \times 10^{-4}\, \mathrm{M}_\odot$\footnotemark[2]\footnotetext[2]{This H mass is chosen to prevent thin-shell instabilities in the donor.} In our models, hot subdwarf stars originating from low-mass progenitors ($< 2.3 \, \mathrm{M_{\odot}}$) have degenerate He cores, leading to He ignition via a core He flash. In contrast, hot subdwarf stars with progenitors  $>$ $2.3 \, \mathrm{M_{\odot}}$ ignite He gradually without undergoing a core He flash. The initial zero-age main sequence (ZAMS) masses of hot subdwarfs are chosen from the range $1\,\mathrm{M}_\odot$ to $5.95\,\mathrm{M}_\odot$ to include both degenerate and non-degenerate core He ignition. The hot subdwarf stars in our models are in the mass range of $0.33 \, \mathrm{M}_{\odot}$ to $0.8 \, \mathrm{M}_{\odot}$. The lower limit of $0.33 \, \mathrm{M}_{\odot}$ represents the minimum mass required for core He ignition. The most massive model, $0.8 \, \mathrm{M}_{\odot}$ expands to a giant-like phase reaching the radii of approximately $10 \, \mathrm{R}_{\odot}$. Notably, our hot subdwarf models include $0.47 \, \mathrm{M}_{\odot}$, which is considered the canonical mass for such stars \citep{2002MNRAS.336..449H,2003A&A...411L.477H}. Table.~\ref{tab:mass_and_periods} gives the details of the donor masses. 

White dwarf: We create carbon oxygen WDs (0.7 to $1.0\,\mathrm{M}_\odot$)  after modifying the \texttt{make\_co\_wd} test suite to cool until the age of $3 \times 10^8\,$ years. The modified test suite \texttt{make\_co\_wd} evolves a single star from the pre-main sequence phase to the WD phase. The mass range of the WD spans from 0.7 to $1.1\, \mathrm{M}_\odot$, chosen to align with the existing observations and to anticipate the potential explosive fate of these systems.  The massive $1.1 \,\mathrm{M}_\odot$ WD is created by rescaling the $1.0 \,\mathrm{M}_\odot$ WD using the MESA inlist parameter \texttt{relax\_mass\_scale}, which rescales to the new mass without altering the composition profile. All the WD models are further refined to convert all H to $^4$He, avoiding the extra computational cost of evolving through classical novae at the onset of mass transfer, which is not the focus of this study. 

Binary Evolution: The mass transfer rate follows MESA's Kolb scheme \citep{1990A&A...236..385K, Paxton2015}. We assume conservative mass transfer for our models. Thus, the mass transfer rate is always equal to the accretion rate. The entropy of the accreted material is determined by matching the entropy of the outer shell of the accretor. Further, the WDs undergoing accretion are expected to undergo compressional heating, which MESA incorporates into evolutionary models through its energy equation and thermodynamically consistent EOS (see \citealt{2023ApJS..265...15J}). Exploring the effects on non-conservative mass transfer, tidal heating, and spin-up of the accretor is left to future work.  

The donor starts transferring the envelope, which adds H and heats the surface of the WD, resulting in classical novae. Simulating and evolving through these novae is computationally expensive and not the focus of this paper. Therefore, we modify the accretion stream to convert the mass fraction of H and $^{3}\text{He}$ to the mass fraction of $^{4}\text{He}$. This treatment is applied consistently throughout the mass transfer phase. Furthermore, the initial composition profile of the WD model is further modified to convert $^{3}\text{He}$ to $^{4}\text{He}$ to avoid further production of H through nuclear reactions.

 We initiate our hot subdwarf and white dwarf models on a regular orbital period grid spanning 36 minutes to 7.2 hours, excluding systems that are Roche-lobe filling at the onset.  This range captures mass transfer at various core He fractions for the donor stars, including phases such as shell burning and thermal pulses. Additionally, the period range allows for donors to evolve without interaction to become a WD. Further details of the grid can be found in the Table \ref{tab:mass_and_periods}. 

The evolution is followed until either the accretor reaches the ignition conditions (see Sect.~\ref{sec:Conditions for ignition}) or the donor evolves to a WD making a double white dwarf. We terminate our simulations as double white dwarfs if the binary system is non-interacting and the radius of the hot subdwarf decreases to $0.03\,\mathrm{R_{\odot}}$. Once this radius is attained, all the hot subdwarf-turned WDs in our simulations enter the cooling track.

We do not include the effects of convective overshooting while creating hot subdwarfs and WDs. All our models assume $Z = 0.02$, with the initial metal fractions taken from \citet{Grevesse1998}. We further assume no gravitational settling, meaning that the hot subdwarf or white dwarf atmosphere contains not only hydrogen but also contamination from heavier elements. Nevertheless, the mass fraction of these elements is likely too low to significantly impact the binary evolution or the final fate of the system.

\begin{table*}
\centering
\begin{tabular*}{\textwidth}{@{\extracolsep{\fill}} c c c}
\hline
Parameter & Range & Values \\
\hline
$\mathrm{M_{hot\ subdwarf}}$ & $0.33 \, \mathrm{M_{\odot}}$ to $0.8 \, \mathrm{M_{\odot}}$ & 0.33, 0.35, 0.39, 0.40, 0.47, 0.50, 0.55, 0.60, 0.65, 0.70, 0.75, 0.80 \\

$\mathrm{M_{WD}}$ & $0.7 \, \mathrm{M_{\odot}}$ to $1.1 \, \mathrm{M_{\odot}}$ & 0.70, 0.75, 0.80, 0.85, 0.90, 0.95, 1.00, 1.10 \\

$\mathrm{P_{i}}$ & 36 minutes to 7.2 hours & Logarithmically spaced (30 values) \\
\hline
\end{tabular*}
\caption{Initial parameters (masses and orbital periods) for hot subdwarfs and WDs. For more details, see Sect.~\ref{sec:Setup}.}
\label{tab:mass_and_periods}
\end{table*}

\begin{figure}
    \centering
     \includegraphics[width=0.5\textwidth]{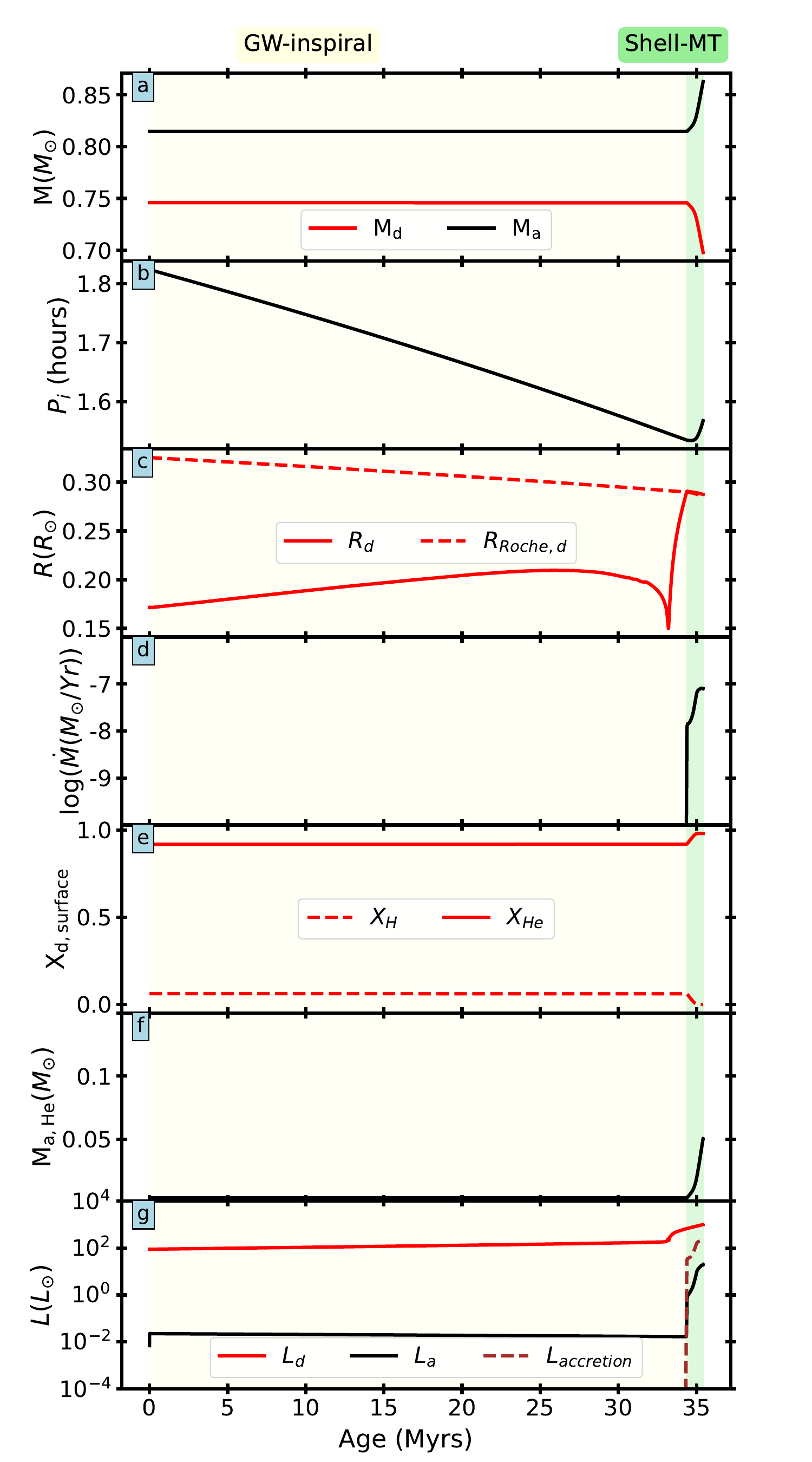}
    \caption{Example of a binary with $\mathrm{M_d = 0.75\, \mathrm{M_{\odot}}}$, $\mathrm{M_a = 0.8\,\mathrm{M_{\odot}}}$, and $\mathrm{P_i = 1.9\, hours}$, where a He nova occurs in the WD (accretor). The panels display the evolution of different parameters as in Fig.~\ref{fig:Explosion}. The background colors yellow, and green represent the gravitational-wave inspiral phase (GW-inspiral), and mass transfer phase during the He shell burning phase (Shell-MT), respectively. The accretor accretes with a relatively high accretion rate of $\sim 10^{-7} \, \mathrm{M_{\odot}yr^{-1}}$, resulting in the ignition in the He layers less dense than the assumed critical density for detonation ($< 10^{6} \, \mathrm{g\,cm^{-3}}$).
}
    \label{fig:Nova}
\end{figure}

\subsection{Conditions for ignition} \label{sec:Conditions for ignition}
 We evolve our binaries either until He ignition occurs in the accretor or when the non-interacting hot subdwarf evolves into a WD. The modeling of white dwarf thermonuclear explosions is inherently a multidimensional problem. \citep{2024A&A...686A.227P}. However, our simulations use hydrostatic 1D MESA calculations, which do not fully capture the multidimensional nature of the explosion or the nature of nuclear burning processes. Therefore, for the binaries in which He ignition occurs in the accretor, we post-process to classify the detonation events. We use an approximate criterion to classify outcomes as either ``detonation'' or ``non-detonation.''
In this classification scheme, we assume all ``detonations'' trigger a second detonation in the core of the WD, leading to a double detonation supernova \citep{2010A&A...514A..53F,2021ApJ...919..126B,2021A&A...649A.155G}. Conversely, ``non-detonations'' result in a He nova, where the He shell is expelled but the WD itself remains intact.
 
From the attempts to understand the distinguishing criteria between detonation and deflagration, previous works have shown that this classification is correlated to various factors such as ignition temperature-dependent critical density \citep{2011ApJ...734...38W}, and density of the ignition region \citep{1994ApJ...423..371W}. From \cite{1994ApJ...423..371W}, the He flashes in the degenerate surface of the WD heavily depend on the density of the ignition region. They found that for the systems undergoing detonation, the minimum density of the He shell was $\rho = 6.8 \times 10^5 \, \text{g\,cm}^{-3}$. Furthermore, most of the systems that experienced detonations were characterized by densities above   $\rho = 10^6 \, \text{g\,cm}^{-3}$. In this work, we adopt a critical density $\rho_c = 10^6 \, \text{g\,cm}^{-3}$. We classify all systems that undergo He ignition at densities below this critical as non-detonations (He novae), and those igniting above this density as detonations leading to thermonuclear supernovae.

\section{Description of example systems} \label{sec:Examples}

In this section, we discuss examples of the three potential outcomes for hot subdwarf + WD binaries as double detonation supernovae, He novae, and double white dwarfs. Fig.~\ref{fig:cartoon} shows a schematic diagram for three different fates. Double detonation supernovae and He novae result from systems that undergo mass transfer and later experience thermonuclear instability in the He layer on the surface of the WD. We differentiate a double detonation supernova from He nova based on the He layer's critical density (see Sect.~\ref{sec:Conditions for ignition}). Double white dwarf systems result from binaries that either do not come into contact or involve episodes of accretion that fail to ignite the He layers of the WD.

\subsection{Supernova} \label{sec:WD explosion}

\begin{figure}
    \centering
    \includegraphics[width=1\columnwidth]{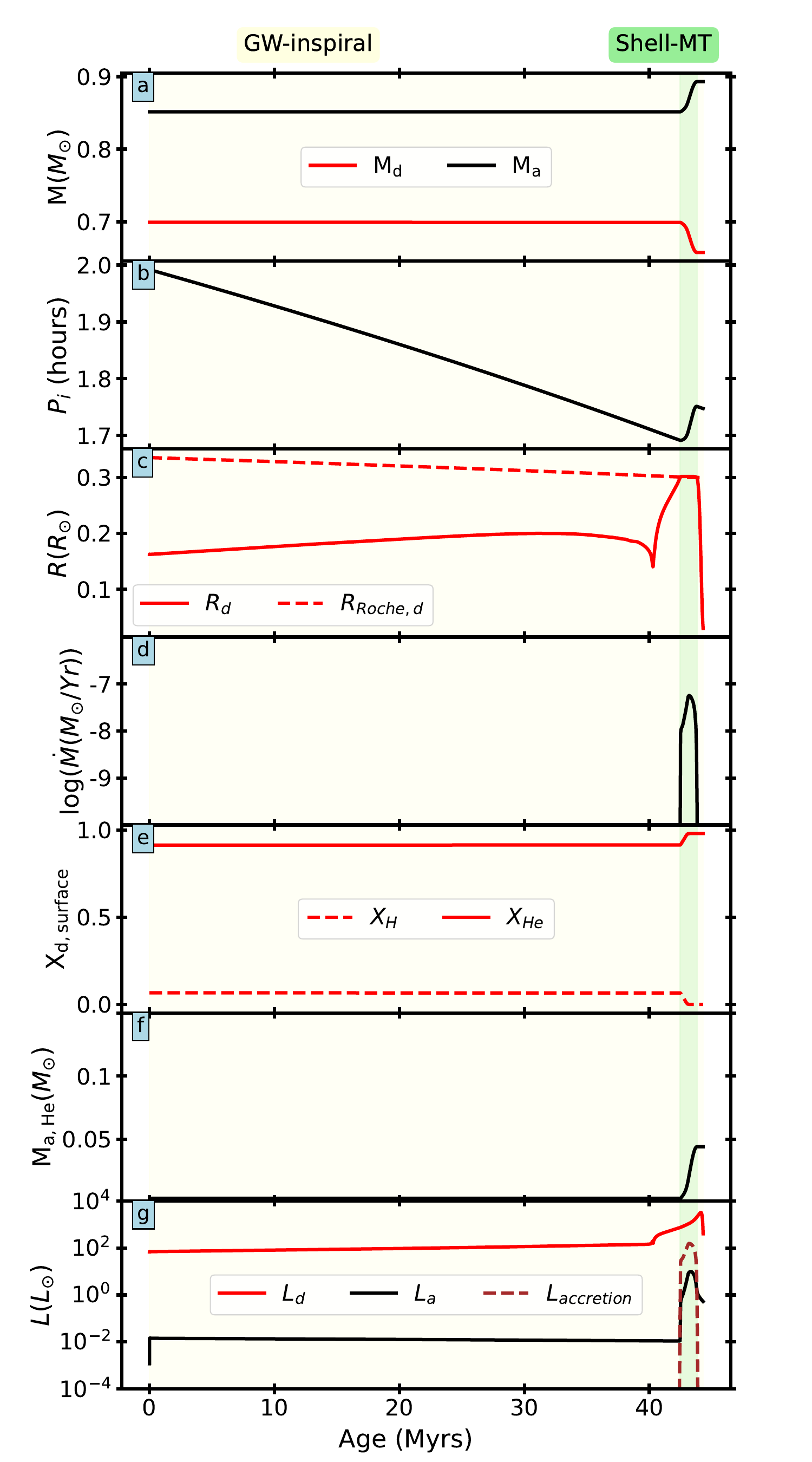}
    \caption{Evolutionary aspects of a binary with $\mathrm{M_d = 0.7 \,\mathrm{M_{\odot}}}$, $\mathrm{M_a = 0.85 \,\mathrm{M_{\odot}}}$, and $\mathrm{P_i = 2\, hours}$, where the donor evolves into a WD, forming a  double white dwarf system. The panels display the evolution of different parameters as  in Fig.~\ref{fig:Explosion}. The background colors yellow, and green represent the gravitational-wave inspiral phase (GW-inspiral), and mass transfer phase during the He shell burning phase (Shell-MT), respectively. This system is detached for most of its lifetime, and undergoes one episode of accretion that fails to ignite the He layer of the accretor, resulting in a double white dwarf. This double white dwarf is expected to merge in 86 million years which might lead to a Type Ia supernova. }
    \label{fig:DWD}
\end{figure}

We first examine a system that evolves into a double detonation supernova. For this example, we initiate a hot subdwarf ($0.5\,\mathrm{M_{\odot}}$) and a WD ($0.8\,\mathrm{M_{\odot}}$) binary system with an orbital period of 0.84 hours, assuming the hot subdwarf has just started core He burning. Figure~\ref{fig:Explosion} illustrates the evolution of key system properties: (a) the mass of the donor and the accretor, (b) the orbital period of the binary, (c) the radius of the donor, (d) the mass transfer rate, (e) the surface mass fractions of H and He in the donor, (f) the He mass in the accretor, and (g) the luminosity of the donor, accretor, and the accretion process which we estimate as accretion luminosity 
\begin{equation}
L_{\text{acc}} \approx \frac{G \, M_{\text{WD}} \, \dot{M}}{R_{\text{WD}}},
\end{equation}
where $ G $ is the gravitational constant, $ M_{\text{WD}} $ is the mass of the WD, $ \dot{M} $ is the mass transfer rate, and $ R_{\text{WD}} $ is the radius of the WD. We estimated $L_{\text{acc}}$ across the entire parameter space, but we did not find a trend that differentiates the outcomes.

\begin{figure*}
    \centering
    \includegraphics[width=\textwidth]{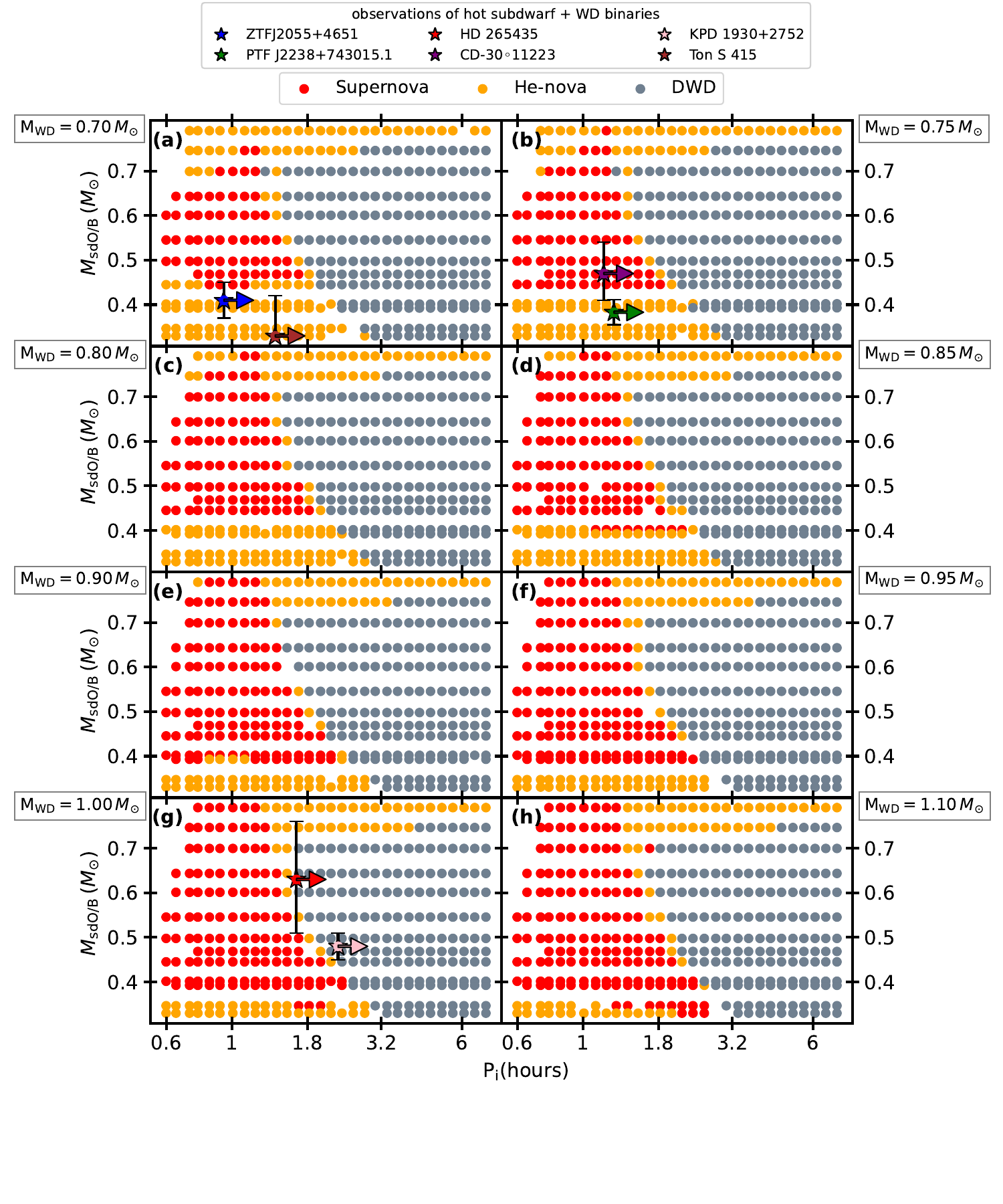}
    \vspace{-80pt}
    \caption{Mapping of outcomes (supernova, He nova, and double white dwarf) across the initial parameter space. The x-axis represents the initial orbital period, and the y-axis represents the initial donor mass ($\mathrm{M_{sdO/B}}$). The 8 sub-panels describe different white dwarf masses, as labeled, that increase from the top left to the bottom right. Red points denote double detonation supernovae, orange points indicate He novae and grey points represent double white dwarfs. The stars highlight observed hot subdwarf + WD binary systems (see Table~\ref{tab:observed_systems}), with arrows indicating that initial orbital periods must be longer than the observed values. The lower mass donors ($<0.4 \, \mathrm{M_{\odot}}$) only lead to double detonation supernovae with higher mass accretors ($>0.9 \, \mathrm{M_{\odot}}$). For details, see Sect.~\ref{sec:Ignition regions across initial parameter space}.}
    \label{fig:grid_analysis}
\end{figure*}

For the initial 11 million years, gravitational wave radiation shortens the binary orbital period until the onset of mass transfer at $\sim 0.55$ hours. In MESA, this inspiral is calculated through the angular momentum loss from gravitational waves, as given by Eq.~4 of \cite{2015ApJS..220...15P} (see also \citealt{1964PhRv..136.1224P}):
\begin{equation}
\dot{J}_{\rm gr} = - \frac{32}{5}\frac{G^{7/3}}{c^5} \left( \frac{2\pi}{P_{\rm orb}} \right)^{7/3} \frac{(M_1 M_2)^2}{(M_1 + M_2)^{2/3}},
\end{equation}
where $M_1$ and $M_2$ are the masses of the two stars, $P_{\rm orb}$ is the orbital period, and $c$ is the speed of light. The hot subdwarf has a convective He burning core and an H-rich envelope. Mass transfer in typical hot subdwarf + WD binaries occurs in two phases: envelope mass transfer and He core mass transfer \citep{2021ApJ...922..245B}. Initially, the donor transfers its H-rich envelope for about 2 million years at a relatively slow mass-transfer rate of about $\sim 10^{-9} \, \mathrm{M_{\odot}yr^{-1}}$. After the envelope is stripped off, the donor transitions to transferring He for the next 7 million years, with the mass-transfer rate increasing to $\sim 10^{-8} \, \mathrm{M_{\odot}yr^{-1}}$(see panel d of  Fig.~\ref{fig:Explosion}), accumulating a dense He layer ($\sim 0.16 \,  \mathrm{M_{\odot}}$) on the surface of the accretor. Throughout the accretion process, the rates are too low for stable He burning and growth of the C/O core, instead He accumulates as the He envelope.  

In the left panel of Fig.~\ref{fig:TRho}, we see the time evolution of the accretor's temperature-density ($T-\rho$) profile. At the start, the temperature decreases monotonically from the center ($\rho \sim 10^7 \, \mathrm{g\,cm}^{-3}$) to the surface. After He accretion starts at about $14 \, \mathrm{Myr}$, a temperature inversion occurs due to compressional heating from the accretion (e.g., \citealt{1982ApJ...253..798N, 2004ApJ...600..390T}). As the accretor continues to accrete He at a steady rate of $\sim 10^{-8} \, \mathrm{M_{\odot}yr^{-1}}$, the combination of compressional heating and temperature diffusion continues to compress and heat the underlying layers, shifting the maximum temperature to deeper He layers with densities $\rho \sim 2 \times 10^{6} \, \mathrm{g\,cm}^{-3}$. This eventually leads to thermonuclear instability in the degenerate He layers, resulting in the runaway fusion of He via the NCO chain (marked as He ignition in Fig.~\ref{fig:TRho}). We stop our simulation at this point.

To understand the dominant process that contributes to reaching the He ignition conditions, we use the analytic expressions described below. We can estimate the adiabatic compression, at constant entropy, using the definition of the third adiabatic index \citep{hansen2004}
\begin{equation}
\left(\frac{\partial \ln T}{\partial \ln \rho}\right)_s \equiv \Gamma_3 - 1 
\end{equation}
\noindent
where $T$ represents the temperature, $\rho$ is the density, and $\Gamma_3$ is the third adiabatic index . The timescale for compressional heating can be estimated as 
\begin{equation}\label{eq:comp_heating}
t_{\text{comp}} = \frac{T}{\dot{T}} = \frac{\rho}{\dot{\rho}} \cdot \left(\frac{1}{\Gamma_3 - 1}\right).
\end{equation}
\noindent
We estimate the heat diffusion timescale using 
\begin{equation}\label{eq:heat_diffusion}
t_{\text{th}} = \frac{H^2}{D_{\text{th}}}
\end{equation}
\noindent
where $H = P/\rho g$ is the local pressure scale height, $P$ is the pressure, and $g$ is the gravitational acceleration. $D_{\text{th}} = 4ac T^3/3\kappa \rho^2 c_P$ is the coefficient of thermal diffusion, where $a$ is the radiation constant, $\kappa$ is the opacity, and $c_P$ is the specific heat at constant pressure.

The neutrino cooling timescale is estimated using 
\begin{equation}\label{eq:neutrino_cooling}
t_{\text{neu}} = \frac{c_P T}{\epsilon_{\text{neu}}}
\end{equation}
\noindent
where $\epsilon_{\text{neu}}$ is the energy loss rate due to neutrino emission. 

For this system, we estimate the timescales for compressional heating, heat diffusion, and neutrino cooling to be approximately $\sim 10^7$ years, $\sim 6 \times 10^5$ years, and $\sim 5 \times 10^6$ years, respectively. Based on the profiles of these quantities from our MESA model, we conclude that heat diffusion is the dominant process in shaping the temperature profile that drives He ignition in the double detonation supernova case described above.

He ignition in this model occurs at densities $\rho \sim 2 \times 10^{6} \, \mathrm{g\,cm}^{-3}$ , which is greater than the adopted critical density (see Sect.~\ref{sec:Conditions for ignition}). Thus we classify this He ignition as a detonation and we expect that triggers a subsequent core detonation in the WD \citep{2019ApJ...873...84P}, consequently destroying the WD in a thermonuclear supernova. Immediately before the detonation, the WD has a mass of $\sim 0.96 \,  \mathrm{M_{\odot}}$ with a thick He shell $\sim 0.16 \,  \mathrm{M_{\odot}}$ and thick He shell likely will lead to a red, faint transient \citep{2010ApJ...719.1067K, 2019ApJ...873...84P}.

\subsection{He nova} \label{sec:He novae}

Next, we focus on a binary system that leads to a He nova. The specific system starts with a $0.75 \,\mathrm{M_{\odot}}$ hot subdwarf donor, a $0.8 \,\mathrm{M_{\odot}}$ WD accretor, and an orbital period of $1.9$ hours. Fig.~\ref{fig:Nova} shows the same key binary evolution properties as shown in the previous section. The stars remain detached throughout the core He burning phase of the hot subdwarf. After about 34 million years, the hot subdwarf exhausts its core He fuel, expands to a He-shell burning phase, and fills its Roche lobe, commencing mass transfer. 
 
 The accretor gains mass at a higher rate ($\sim 10^{-7} \, \mathrm{M_{\odot}yr^{-1}}$) compared to the double detonation supernova example, resulting in the transfer of $0.05 \,\mathrm{M_{\odot}}$  of He over 1 million years. We note that the mass transfer rates in this scenario are about an order of magnitude larger compared to the supernova example. At these higher accretion rates, the timescale for compressional heating is estimated to be $\sim 124$ years (Eq.~\ref{eq:comp_heating}), which is shorter than the estimated timescale for heat diffusion, $\sim 0.2 \times 10^5$ years (Eq.~\ref{eq:heat_diffusion}).
 Thus, the accretion compresses and heats the less dense layers, eventually triggering a thermonuclear instability. This leads to unstable He burning  via the $3\alpha$ process. We stop our simulations at this point.

The right panel of Fig.~\ref{fig:TRho} depicts the evolution of the $T-\rho$ profile of the accretor and marked regions show the He ignition. Since the He ignition occurs at densities $\rho \sim  2.5 \times 10^{5} \, \mathrm{g\,cm}^{-3}$, which is less than the critical density (see Sect.~\ref{sec:Conditions for ignition}), we expect that this system does not create a detonation when it ignites. While we do not evolve systems beyond the initial thermonuclear instability, previous studies have found various outcomes for the systems undergoing He novae. Some studies \citep{2008AstL...34..620Y,2010Ap&SS.329...25N,2015ApJ...807...74B} have found that for systems that survive the initial thermonuclear instability, subsequent evolution is unlikely to lead to a supernova via double detonation. Instead, the continued mass transfer will lead to weaker He shell flashes. However, some studies \citep{2001ApJ...558..323H,2016A&A...589A..43N} have found evidence for He-novae being progenitors of Type Ia supernova. Thus, the deterministic fate of these systems can be stated only after full evolution through all the nova outbursts.

\subsection{Double white dwarf} \label{sec:DWD}

\begin{figure*}
    \centering
    \includegraphics[trim = 0 0 0 0 ,width=1\textwidth]{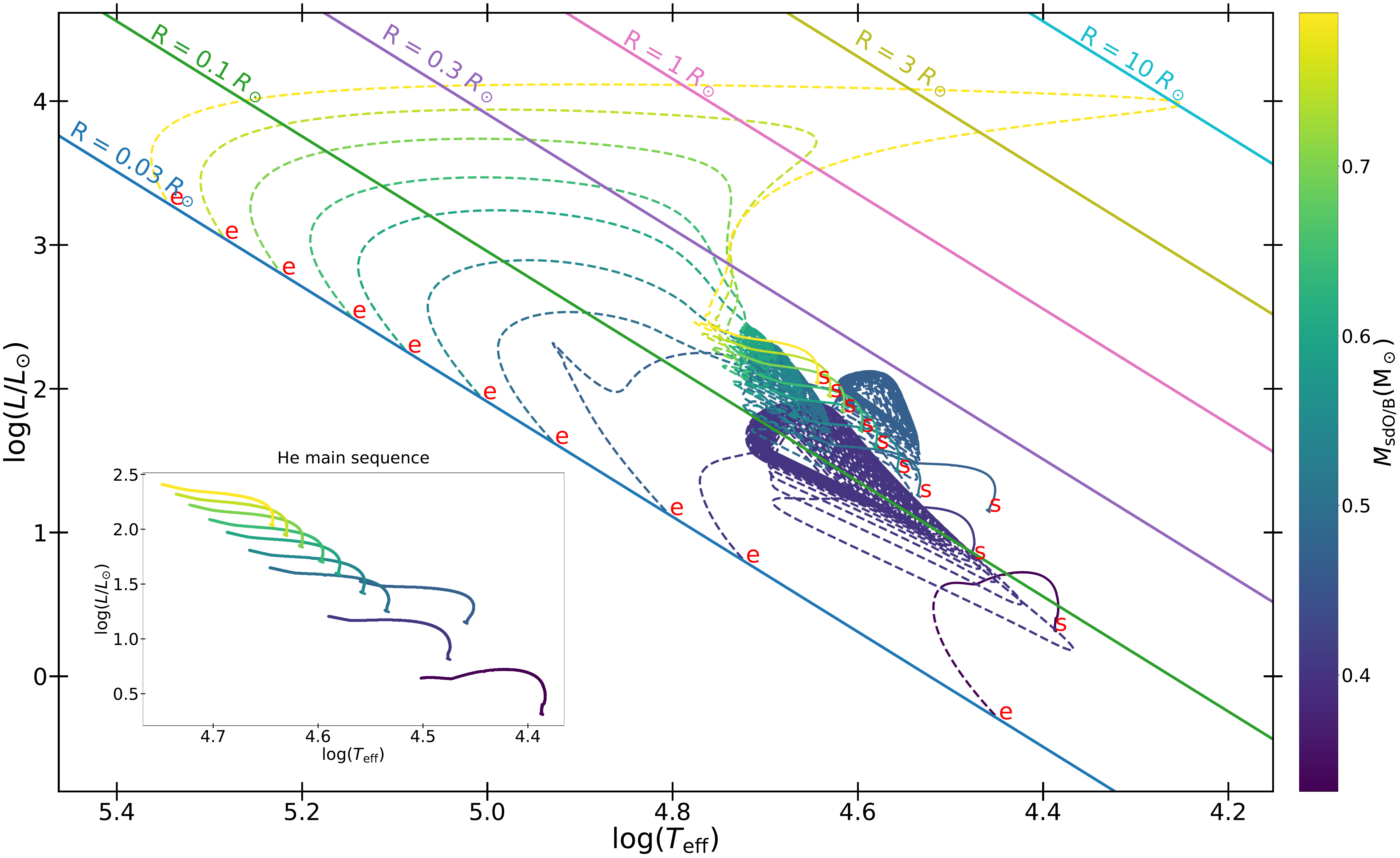}
    \caption{Evolution of selected single hot subdwarf stars is shown in the HR diagram. The annotated red letters ``s" and ``e" indicate the start and end of the evolution respectively. Different colors represent different masses of hot subdwarf stars, as indicated by the color bar. The constant radius lines illustrate the radius evolution of the hot subdwarf stars during their He shell burning phase. The inset represents a zoomed-in view of the He main sequence phase. Details of the evolution of hot subdwarf as single stars are explained in Sect.~\ref{sec:Setup}}
    \label{fig:HR_donor}
\end{figure*}

Finally, we look at a system that does not ignite explosively and evolves into a double white dwarf binary. Fig.~\ref{fig:DWD} shows the binary evolution properties of the system. The system starts as a hot subdwarf ($0.7 \, \mathrm{M_{\odot}}$) and a WD ($0.85 \, \mathrm{M_{\odot}}$) binary system with an orbital period of 2 hours, assuming the hot subdwarf has just started core He burning. The system comes into contact during the giant-like phase of the hot subdwarf after evolving without interaction for about 42 million years.

The system undergoes one episode of accretion  with a rate of $\sim 10^{-7} \, \mathrm{M_{\odot}yr^{-1}}$ depositing $\sim 0.04 \, \mathrm{M_{\odot}}$ of He on the surface of the accretor. However, the mass transfer stops afterward as the donor runs out of He and evolves into a WD, resulting in a double white dwarf. The accretor (originally a WD) in this double white dwarf has a massive shell due to accretion, unlike those formed through single stellar evolution. We terminate our simulation when the donor's radius decreases to $0.03 \, \mathrm{R_{\odot}}$ and the system is no longer interacting (for more details, see Sect.~\ref{sec:Setup}). As seen in panel (b) of Fig.~\ref{fig:DWD}, the orbital period increases due to mass transfer. However, the magnitude of this increase is not very significant compared to the overall shortening from the initial orbital period (2 hours). The resulting double white dwarf has masses of  $\sim 0.66 \, \mathrm{M_{\odot}}$ and $ \sim 0.89 \, \mathrm{M_{\odot}}$, with an orbital period of 1.75 hours. This double white dwarf is expected to merge in 86 million years. Given that one of the WDs has a mass of $\sim 0.9 \, \mathrm{M_{\odot}}$, the example system could potentially lead to a Type Ia supernova triggered by the merger \citep{2021MNRAS.503.4734P}.

\section{The fate of binaries across our parameter space} \label{sec:Ignition regions across initial parameter space}

We construct a grid of binaries varying in initial accretor masses, orbital periods, and donor masses. The ranges of donor mass ($\mathrm{M_{hot subdwarf}}$), accretor mass ($\mathrm{M_{WD}}$), and initial orbital periods ($\mathrm{P_{i}}$) are listed in Table~\ref{tab:mass_and_periods} (see also Sect.~\ref{sec:Setup}). Fig.~\ref{fig:grid_analysis} presents the main results, showing the fates of the systems as a function of their initial parameters. Each point represents the outcome of a MESA model of a hot subdwarf + WD binary. Eight panels correspond to different initial WD masses. Within each panel, we vary the initial masses for the donor and different initial orbital periods. Systems are color-coded based on their outcomes: red points indicate systems that end in double detonation supernovae, orange points represent systems that lead to He novae, and grey points denote systems that evolve into double white dwarfs. The stars of different colors mark observed systems within our parameter space (see Table~\ref{tab:observed_systems}, \citealt{2000MNRAS.317L..41M,Geier2007,Geier2013,2020ApJ...891...45K,2021NatAs...5.1052P,2023MNRAS.525..183S,2024MNRAS.527.2072D}). In the following subsections, we discuss the donor's impact on the mass transfer rate, how the accretor affects the mass transfer rate. Lastly, we discuss the two interesting outliers.
\begin{table*}
\centering
\begin{tabular*}{\textwidth}{@{\extracolsep{\fill}} l 
                               c
                               c 
                               c 
                               l}
\hline
Observed system & Orbital Period (min) & $M_{\text{hot subdwarf}} \, (\mathrm{M_{\odot}})$ & $M_{\text{WD}} \, (\mathrm{M_{\odot}})$ & Reference \\ 
\hline
ZTFJ2055+4651  & $56.34785 \pm 0.00026$ & $0.41 \pm 0.04$ & $0.68 \pm 0.05$ & \citet{2020ApJ...898L..25K} \\ 
PTF J2238+743015.1  & $76.341750 \pm 0.000001$ & $0.383 \pm 0.028$ & $0.725 \pm 0.026$ & \citet{2022ApJ...925L..12K} \\ 
HD 265435  & $99.09918 \pm 0.00029$ & $0.63^{+0.13}_{-0.12}$ & $1.01 \pm 0.15$ & \citet{2021NatAs...5.1052P} \\ 
CD-30$^\circ$11223  & 70.53 & $0.47^{+0.07}_{-0.06}$ & $0.74 \pm 0.02$ & \citet{Geier2013,2024MNRAS.527.2072D} \\ 
KPD 1930+2752  & 137 & $0.48 \pm 0.03$ & $1.0 \pm 0.03$ & \citet{2000MNRAS.317L..41M,Geier2007} \\ 
Ton S 415  & $84.646 \pm 0.0004$ & $0.33 \pm 0.09$ & $0.47 \pm 0.24$ & \citet{2023MNRAS.525..183S} \\ 
\hline
\end{tabular*}
\caption{Observed systems with their orbital periods, hot subdwarf masses, and WD masses.}
\label{tab:observed_systems}
\end{table*}

\subsection{Effect of the donor mass on the mass transfer rate}\label{Effect of donor on the mass transfer rate}

\begin{figure*}
    \centering
    \includegraphics[trim = 0 0 0 0 ,width=1\textwidth]{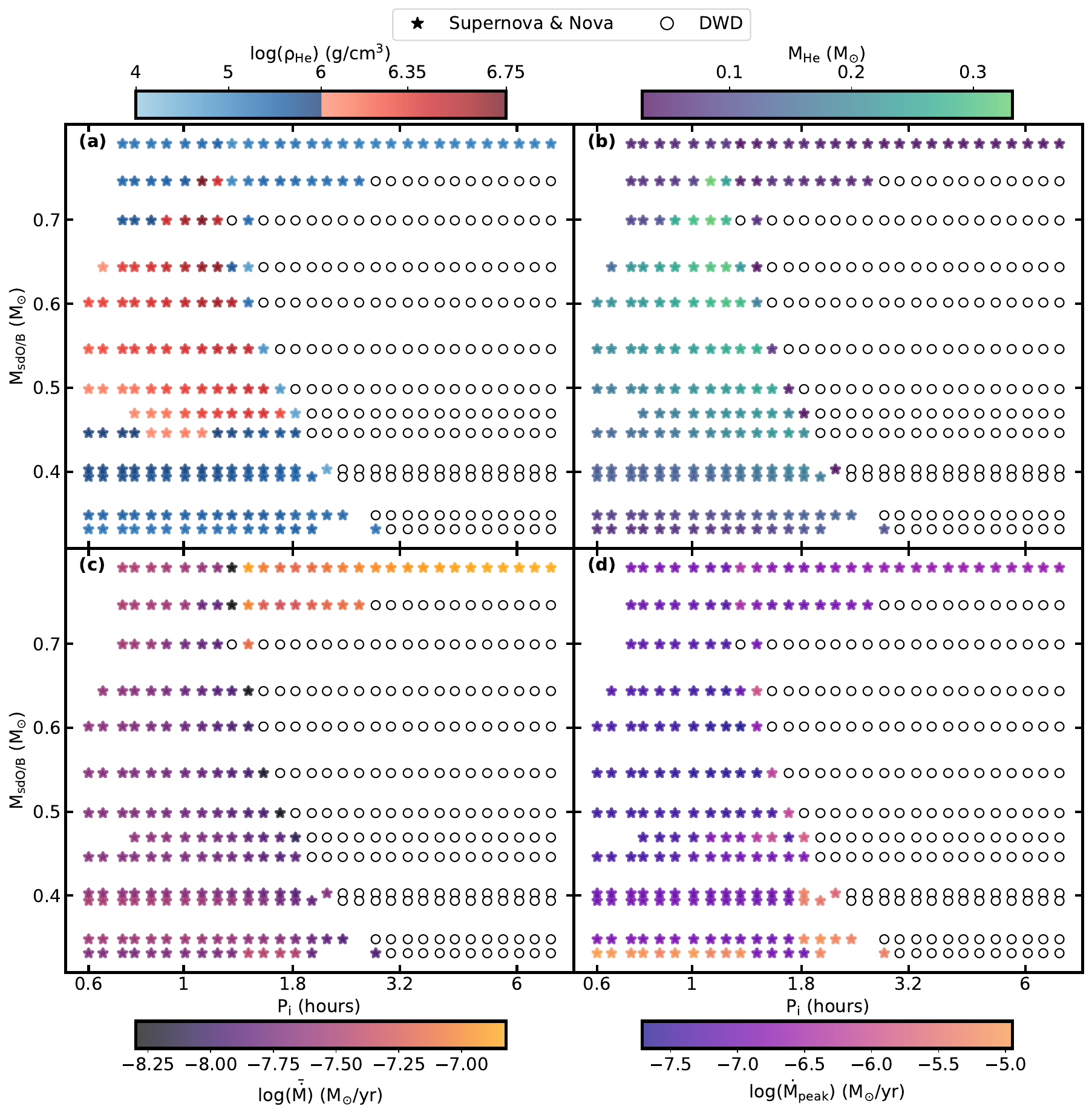}
    \caption{Critical properties of systems that undergo He ignition in binaries with an accretor of an initial mass of $0.7\,\mathrm{M_{\odot}}$. The figure shows (a) the density of the He layer at the location of ignition, (b) the total mass of He accumulated on the WD, (c) the time-averaged accretion rate from the time when the accretion rate was greater than $\mathrm{10^{-10}}\,\mathrm{M_{\odot}\,yr^{-1}}$ until the He ignition and (d) the peak accretion rate at which the WD gains mass. Systems classified as double detonation supernovae experience steady accretion with lower average and peak rates, leading to the accumulation of more He and ignition at a denser ($> 10^{6} \, \mathrm{g \, \text{cm}^{-3}}$) layer.}
    \label{fig:0.7_ignition properties}
\end{figure*}

In this section, we present the different evolutionary phases of hot subdwarfs and how they vary with the initial mass of the hot subdwarf. This difference is crucial for understanding the results of binary evolution, as it is one of the causes that determine the point of contact during mass transfer.

Fig.~\ref{fig:HR_donor} illustrates the HR diagram for the evolution of hot subdwarf stars in isolation. Low-mass hot subdwarf stars ($< 0.6 \, \mathrm{M_{\odot}}$) begin by burning He in the core, enter a thermally pulsing phase once core He is exhausted, and eventually evolve into WDs with thick He shells. In our MESA models, these stars experience periodic thermal pulses because the He shell burning is unstable. 
The pulses cause the star to undergo significant expansions and contractions, leading to variability in the luminosity. Conversely, high-mass hot subdwarf stars ($> 0.6 \, \mathrm{M_{\odot}}$) start with core He burning and transition to stable He shell burning upon exhausting core He. After the shell-burning phase, these stars also become WDs with thick He shells. 

For accretor and donor masses $< 0.75 \,\mathrm{M_{\odot}}$, we observe an average accretion rate of approximately $10^{-8} \,\mathrm{M_{\odot} \,yr^{-1}}$ during the core He burning phase of the donor (see panel (c) of Fig.~\ref{fig:0.7_ignition properties}). During the He shell burning, thermally pulsing phase, and proto-WD (where donor's luminosity comes from residual heat rather than fusion) phase, the peak mass transfer rate ranges from about $10^{-7} \,\mathrm{M_{\odot} \,yr^{-1}}$ to $10^{-6} \,\mathrm{M_{\odot}\, yr^{-1}}$ (see panel (d) of Fig.~\ref{fig:0.7_ignition properties}). If mass transfer starts during the core He burning phase, the accretion rates are generally slow enough to allow accumulation of large He shells that result in He ignition in the deeper, denser ($\mathrm{\rho > 10^{6}\, g\,cm^{-3}}$) (see panel (a) of Fig.~\ref{fig:0.7_ignition properties})  layers of the accretor, resulting in a double detonation supernova. Conversely, if mass transfer occurs during the He shell burning, thermally pulsing phase, or proto-WD phase of the donor, He ignition occurs in the less dense ($\mathrm{\rho < 10^{6}\, g\,cm^{-3}}$) (see panel (a) of Fig.~\ref{fig:0.7_ignition properties})  layers of the accretor, leading to a He nova. For donor masses $> 0.75 \, \mathrm{M_{\odot}}$, where there is the possibility of the donor mass being greater than that of the accretor in low mass accretors (for example, a $0.7 \, \mathrm{M_{\odot}}$ WD and a $0.75 \, \mathrm{M_{\odot}}$ WD), the mass transfer rate depends on the mass of the accretor (for more details, see Sect.~\ref{Effect of accretor on the mass transfer}).In our simulations, mass transfer ceases only when the donor exhausts its He or when we stop the simulation when it reaches one of the specified outcomes. 

We use systems containing 0.7 WD as examples and explain them in detail. Panel (a) of Fig.~\ref{fig:grid_analysis} displays outcomes from a grid of binary systems featuring a $0.7 \, \mathrm{M_{\odot}}$ accretor and various donors at different initial masses and binary orbital periods. Fig.~\ref{fig:0.7_ignition properties} displays the critical properties of the systems that undergo He ignition in binaries with an accretor of an initial mass of $0.7\,\mathrm{M_{\odot}}$. The figure illustrates (a) the density of the He layer at the location of ignition, (b) the total mass of He accumulated on the WD, (c) the time-averaged accretion rate for times when the instantaneous accretion rate was greater than $\mathrm{10^{-10}}\,\mathrm{M_{\odot}\,yr^{-1}}$, and (d) the peak accretion rate at which the WD gains mass. 

First, we describe systems containing low-mass hot subdwarf donors (ranging from $0.33 \, \mathrm{M_{\odot}}$ to $0.40 \, \mathrm{M_{\odot}}$). For these systems, He ignition occurs in WDs when the initial orbital periods are less than 3 hours. In contrast, no He ignition occurs for initial orbital periods longer than 3 hours. Of the binaries where the accretor undergoes He ignition, the system comes into contact and starts transferring mass during the core He burning phase of the donor and continues as the donor evolves into a WD. Throughout the core He burning phase, they accrete at the rate of $\sim 10^{-8} \, \mathrm{M_{\odot} \, \text{yr}^{-1}}$. When the donor transitions into a proto-WD, the mass transfer rates peak at $\sim 10^{-6} \, \mathrm{M_{\odot} \, \text{yr}^{-1}}$, resulting in He ignition in less dense ($< 10^{6} \, \mathrm{g \, \text{cm}^{-3}}$) layers of the WD. This ignition occurs at densities less than the critical density (see Sect.~\ref{sec:Conditions for ignition}), classifying all of them as He novae. Systems with low-mass donors ($0.33 \, \mathrm{M_{\odot}}$ to $0.40 \, \mathrm{M_{\odot}}$) and initial orbital periods longer than $\sim 3$ hours never interact before the hot subdwarf becomes a WD. The initial orbital periods of the double white dwarf systems range from 3 to 7.2 hours. These systems are expected to merge due to gravitational wave radiation in a time ranging from 81 million years to 7.9 billion years. The time until a binary system merges due to gravitational wave radiation \citep{1964PhRv..136.1224P} is calculated by

\begin{equation}
t_{\text{merge}} = \frac{5}{256} \frac{c^5}{G^3} \frac{a^4}{M_a M_d (M_a + M_d)}
\end{equation}
\noindent
where and $a$ denotes the semi-major axis of the orbit.

For slightly more massive donors than described in the previous paragraph ($0.45 \, \mathrm{M_{\odot}} \leq \mathrm{M_d} \leq 0.6 \, \mathrm{M_{\odot}}$), He ignition occurs for initial orbital periods less than 2 hours. For initial orbital periods longer than 2 hours, the binary system fails to ignite the He layers of the WD. For systems that experience He ignition, donors in this mass range initiate mass transfer either during the core He burning stage or during the thermally pulsing phase, depending on the initial orbital period. If mass transfer begins during the core He burning phase and proceeds with steady accretion at a rate of approximately $10^{-8} \, \mathrm{M_{\odot} \, \text{yr}^{-1}}$, it can lead to He ignition in layers of the WD where densities exceed the critical density criterion, leading to classification as a double detonation supernova. Conversely, if mass transfer begins during the core He burning phase and continues as the donor transitions to a WD, or if it starts during the thermally pulsing phase, the peak accretion rates range from $\sim 10^{-7} \, \mathrm{M_{\odot} \, \text{yr}^{-1}}$ to $\sim 10^{-6} \, \mathrm{M_{\odot} \, \text{yr}^{-1}}$. This results in He ignition in the less dense layers of the WD, leading to these systems being classified as He novae. For initial orbital periods between 2 and 3 hours, the WD experiences episodic accretion during the thermally pulsing phase of the donor before the donor exhausts its He. However, this accretion does not lead to He ignition, resulting in the formation of a double white dwarf system. For systems with donor masses in the range of $0.45 \, \mathrm{M_{\odot}} \leq \mathrm{M_d} \leq 0.6 \, \mathrm{M_{\odot}}$ and initial orbital periods ranging from 3 to 7.2 hours, double white dwarfs are expected to merge within 3.8 million years to 6.1 billion years.

For more massive donors ($0.6 \, \mathrm{M_{\odot}} \leq \mathrm{M_d} \leq 0.75 \, \mathrm{M_{\odot}}$), He ignition in the accretor occurs for shorter initial orbital periods (less than 1.5 hours). For initial orbital periods longer than 1.5 hours, the binary system fails to ignite the WD. For systems that undergo ignition, if the mass transfer commences during the core He burning phase, the WD gains mass at a steady rate of $\sim 10^{-8} \, \mathrm{M_{\odot} \, \text{yr}^{-1}}$, leading to ignition in He layers with densities greater than the critical density criteria, classifying these systems as double detonation supernovae. For systems undergoing mass transfer during the shell He burning phase, or during the core He burning phase that continues into the shell He burning phase, the peak accretion rate ranges from $\sim 10^{-7} \, \mathrm{M_{\odot} \, \text{yr}^{-1}}$ to $\sim 10^{-6} \, \mathrm{M_{\odot} \, \text{yr}^{-1}}$. This results in ignition in the less dense He layers of the WD, classifying these systems as He novae. For donor masses between $0.6 \, \mathrm{M_{\odot}}$ and $0.7 \, \mathrm{M_{\odot}}$ and initial orbital periods of 1.5 to 2.5 hours, the system undergoes episodes of accretion during the shell burning phase of the donor before the donor exhausts He. For initial orbital periods greater than 2.5 hours, the donor and accretor do not interact and evolve in isolation to become a double white dwarf. These double white dwarfs will eventually merge in approximately 2 million years to 4.8 billion years.

A $0.8 \, \mathrm{M_{\odot}}$ donor whose mass is higher than the accretor ($q>1; q = \mathrm{M_a}/\mathrm{M_d}$) transfers mass at a high average accretion rate ($\sim 10^{-7} \, \mathrm{M_{\odot} \, \text{yr}^{-1}}$) at all initial orbital periods within our range. This high accretion rate ignites the less dense outer layers of the white dwarf. However, the rates are not sufficient to sustain steady helium burning and subsequent C/O core growth. We therefore classify this system as a He nova. Our initial orbital period range is insufficient for these systems to evolve in isolation to become double white dwarfs.

As the donor mass increases, the rate of He burning also increases, leading to a shorter core He burning phase. During this phase, these binaries come into contact after losing angular momentum due to gravitational wave radiation. Because of the shorter core He burning phase at higher masses, there is less time for gravitational wave radiation to shrink the orbit. Consequently, the range of initial orbital periods that result in double detonation supernovae becomes smaller with higher donor masses, while the range for double white dwarfs becomes larger.

\begin{figure*}
    \centering
    \includegraphics[trim = 0 0 0 0 ,width=1\textwidth]{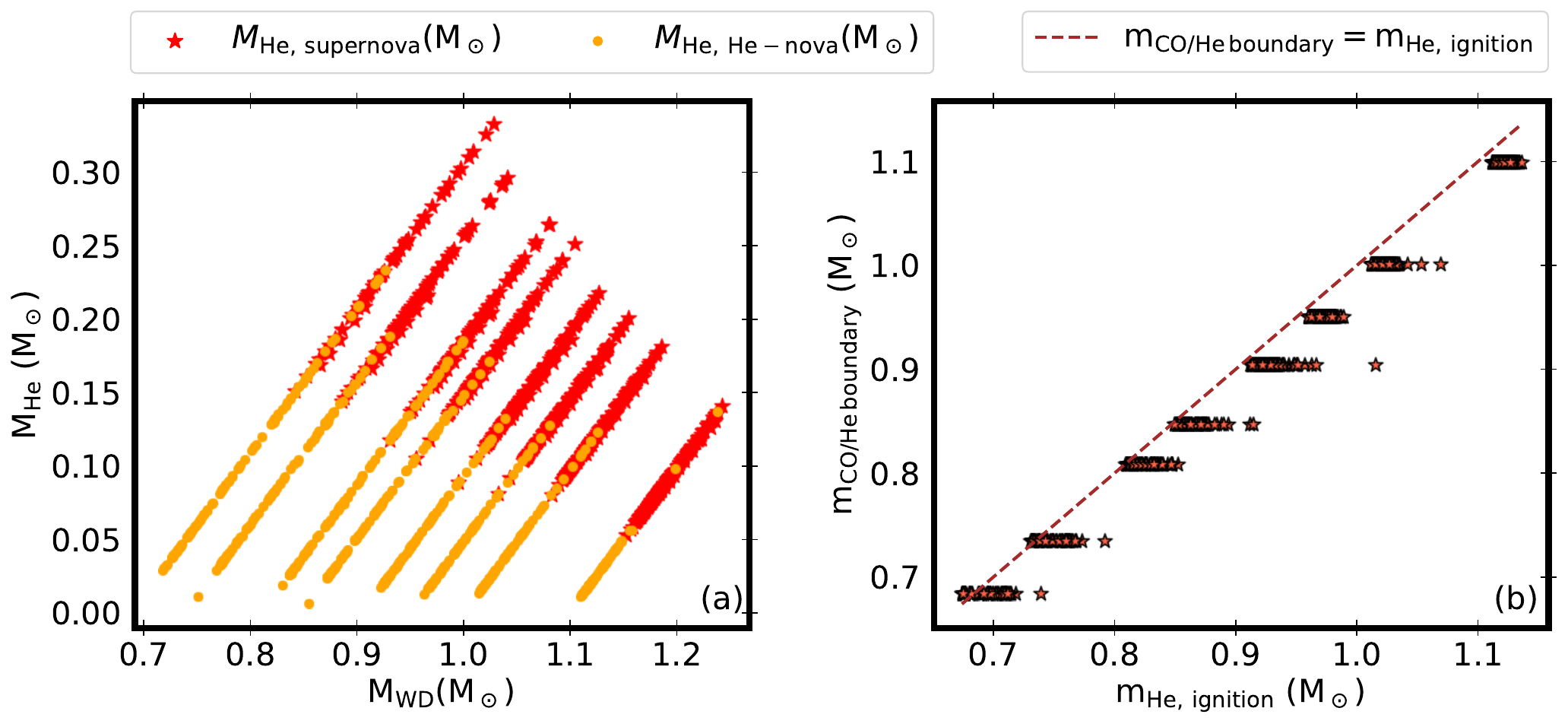}
    \caption{He masses (panel a) and mass coordinates (panel b) for systems that undergo double detonation supernovae and He novae. Panel (a) shows the mass of the He shell at the time of ignition versus the total mass of the WD for systems undergoing double detonation supernovae and He novae. The minimum He mass required for He ignition in our simulations, at  high densities greater than the assumed critical density is $\sim 0.05\,\mathrm{M_\odot}$. Panel (b): Red stars with black edges display the mass coordinate of the ignition point versus the mass coordinate of the C/O--He boundary, where the C/O-He boundary is defined as the layer containing less than $10\%$ He. The red dashed line represents the locus where the mass coordinate of the C/O-He boundary equals that of the ignition point.}
    \label{fig:Explosion_Heshell}
\end{figure*}

\subsection{Effect of the initial accretor mass on the accretion outcome}\label{Effect of accretor on the mass transfer}

The initial accretor mass can affect the fate of the binary for two reasons. Firstly, massive WDs have a denser He shell compared to low-mass WDs.
For instance, the base of the He shell prior to accretion in our $1.0\, \, \mathrm{M_{\odot}}$ WD is denser ($ \rho \sim 2 \times 10^5 \, \mathrm{g \, cm^{-3}}$) compared to that in a $0.7 \, \mathrm{M_{\odot}}$ WD ($ \rho \sim 0.8 \times 10^5  \, \mathrm{g \, cm^{-3}}$). He ignition is highly dependent on both temperature and density. As a result, the mass of the He required to ignite the  $1.0\,\mathrm{M_{\odot}}$ accretor is significantly lower than that required for the $0.7\,\mathrm{M_{\odot}}$ accretor. As a result, for a given accretion rate, the time needed to ignite He in the $1.0\,\mathrm{M_{\odot}}$ accretor is shorter compared to the $0.7\,\mathrm{M_{\odot}}$ accretor. This effect is explained below by comparing binaries with two different accretors $0.7\,\mathrm{M_{\odot}}$ and $1.0\,\mathrm{M_{\odot}}$ but with the same donor mass $0.4 \, \mathrm{M_{\odot}}$ and same initial orbital period of 36 minutes.

With a low-mass donor ($0.4 \, \mathrm{M_{\odot}}$) and an initial orbital period of 36 minutes, the $1.0 \, \mathrm{M_{\odot}}$ WD gains mass during the donor's core He burning phase at an average mass transfer rate of $10^{-8} \, \mathrm{M_{\odot} \, yr^{-1}}$, accumulating approximately $0.09 \, \mathrm{M_{\odot}}$, which is sufficient to trigger He ignition. The density of the He ignition is greater than the critical density, making the outcome for this system classified as a double detonation supernova. However, with the same donor and initial orbital period, a $0.7 \, \mathrm{M_{\odot}}$ WD accretor begins to gain mass during the donor's core He burning phase but accumulates about $0.11 \, \mathrm{M_{\odot}}$ at the similar average rate. This is insufficient to heat the He layers for ignition, given the lower density of the He shell in the $0.7 \, \mathrm{M_{\odot}}$ WD compared to the $1.0 \, \mathrm{M_{\odot}}$ WD. As a result, mass transfer continues as the donor transitions into a WD. During this transition, the mass transfer rate increases to around $10^{-7} \, \mathrm{M_{\odot} \, yr^{-1}}$, accumulating a total of approximately $0.12 \, \mathrm{M_{\odot}}$, leading to He ignition in the less dense layers of the $0.7 \, \mathrm{M_{\odot}}$ WD. The density of the He ignition is less dense than the critical density. Hence we classify these systems as He novae. 

Secondly, the mass ratio has a direct effect on the mass transfer rate. The mass ratio affects the response of the Roche radius to the mass transfer and hence the mass accretion rate. \cite{2021ApJ...922..245B} studied the phases of mass transfer in hot subdwarf + WD binaries. In particular, they present an analytic expression to estimate the Roche radius response to the mass transfer as $({\mathrm{d} \log R_{\text{RL}}}/{\mathrm{d} \log M_{d}}) \approx - \frac{5}{3} + 2.1 q$. Let us compare binaries with two different initial accretor masses, $0.7 \, \mathrm{M_{\odot}}$ and $1.0 \, \mathrm{M_{\odot}}$, but with the same donor of $0.8 \, \mathrm{M_{\odot}}$ and same initial period of 36 minutes. For an accretor of $0.7 \, \mathrm{M_{\odot}}$ and a donor of $0.8 \, \mathrm{M_{\odot}}$, the mass ratio $q$ is greater than 1, but for a $1.0 \, \mathrm{M_{\odot}}$ accretor, $q < 1$.  For a larger mass ratio as in  $0.7 \, \mathrm{M_{\odot}}$ accretor and a $0.8 \, \mathrm{M_{\odot}}$ donor, the Roche lobe contracts as the donor loses mass, resulting in an increased mass transfer rate. Whereas for a $1 \, \mathrm{M_{\odot}}$ accretor, the Roche radius of the donor increases in response to mass loss, leading to a relatively smaller accretion rate, leading to a double detonation supernova.

\subsection{Outliers}\label{sec:outliers}
In this section, we present two interesting outlier systems that deviate from the overall trends of the grid in Fig.~\ref{fig:grid_analysis}. The detailed evolution diagram can be found in appendix~\ref{Appendix:app_ouliers}.

\subsubsection{Double white dwarf with massive He shell}\label{subsec:DWD with massive He shell}

We explain how a $0.7\,\mathrm{M}{_\odot}$ accretor accumulates $\sim0.18\, \mathrm{M{_\odot}}$ of He. The binary system, comprising a $0.7\,\mathrm{M}{_\odot}$ donor, a $0.7\,\mathrm{M}{_\odot}$ accretor, and an initial orbital period of $\sim 1.3$ hours, undergoes two episodes of accretion. The first mass transfer episode begins during the donor's core He burning phase, which heats the accretor, but pauses when the donor contracts due to core He depletion, allowing the accretor to cool. The second mass transfer episode occurs as the donor expands during its shell-burning phase, leading to the total accumulation of $\sim 0.18\, \mathrm{M_{\odot}}$ of He in the accretor.

\subsubsection{Supernova during the late giant phases of accretion} \label{subsec:wide period supernova}

We explain the formation of a double detonation supernova during shell burning phase of a donor. A binary with a $0.7\, \mathrm{M}_{\odot}$ donor, a $1.1\, \mathrm{M}_{\odot}$ accretor, and an initial orbital period of $\sim 1.7\,$hours comes into contact during the late shell burning phase of the donor. Unlike He nova systems with shorter orbital periods around 1.5 hours (where accretion rates reach about $10^{-7}\, \mathrm{M_{\odot}\,yr^{-1}}$), this system begins interacting during a later expansion phase, where the donor's expansion rate is slower, resulting in lower accretion rates of around $10^{-8}\, \mathrm{M_{\odot}\,yr^{-1}}$. This lower accretion rate leads to He ignition in the dense layers of the WD, ultimately getting classified as a double detonation supernova.

\section{Final state of the binaries and Observables}\label{sec:Ignition properties}

In this section, we present the final state of the binaries from our simulations, including the mass of the He shell of the WD at the time of the explosion and the orbital velocity at time of the explosion that should be comparable to the runaway velocity of the surviving star. Additionally, we present the mass of He shell of both the donor and the accretor for systems that end up as double white dwarfs.

\subsection{He shell masses at ignition} \label{sec:M_He}

\begin{figure}
    \centering
    \includegraphics[width=1\columnwidth]{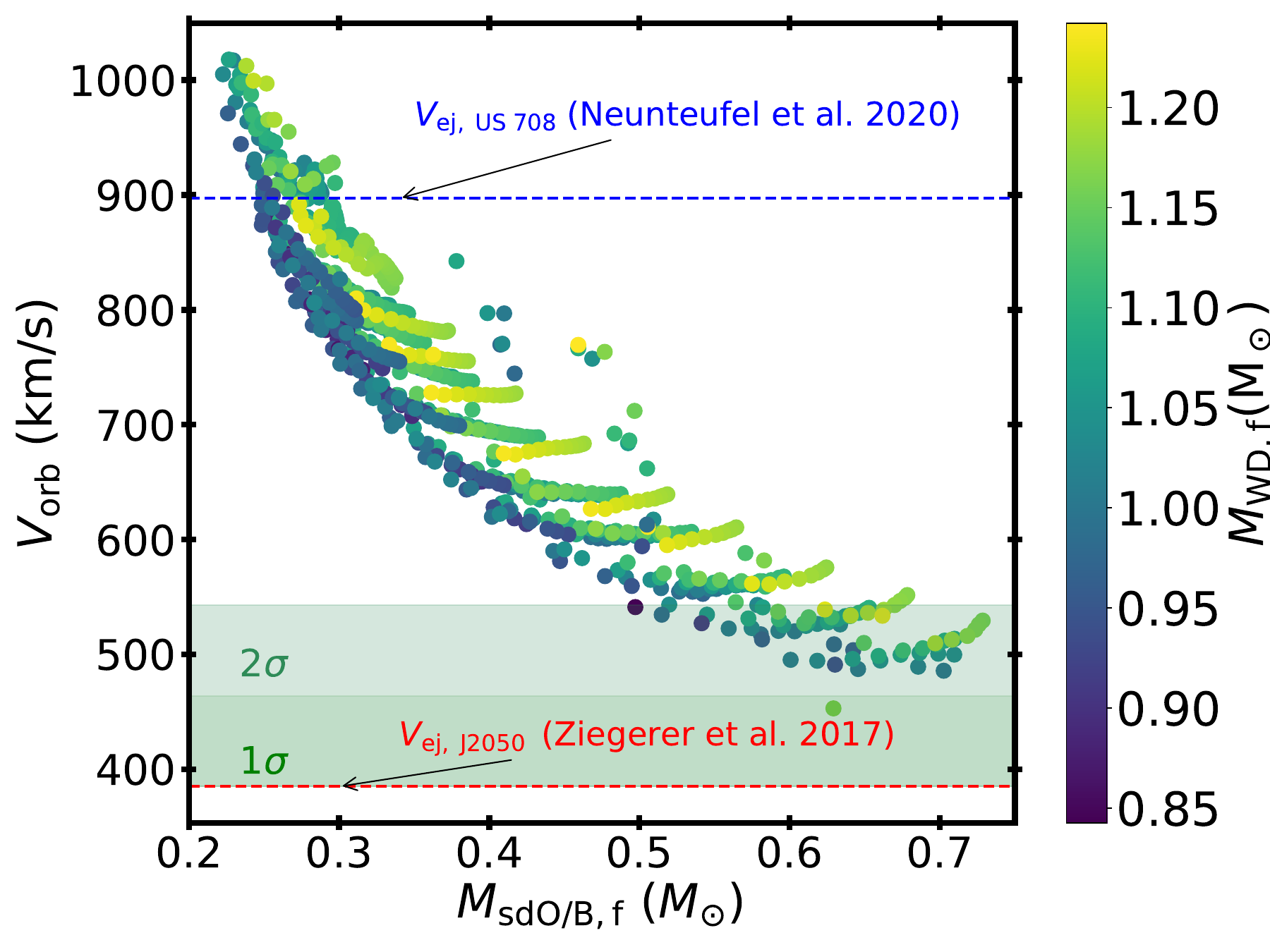}
    \caption{Orbital velocities of the donor at the time of the explosion of the WD. The blue line and red line indicate inferred ejection velocities of the observed He-rich stars US708 (\citealt{2020A&A...641A..52N}) and J2050 (\citealt{2017A&A...601A..58Z}) respectively. The dark and light green shaded regions represent the $1\,\sigma$ and $2\,\sigma$ uncertainties to J2050. Low-mass donors with high-mass accretors result in the highest possible orbital velocities. The outliers with higher velocities for the same initial donor mass indicate the donor is a  proto-WD at the time of He ignition.}
    \label{fig:v_orb}
\end{figure}

He-accreting WDs are one of the progenitors of double detonation supernovae. Sub-Chandrasekhar explosions can occur when explosive burning in the He shell of the WDs triggers a second detonation within the WD, resulting in a thermonuclear supernova. However, the He mass required for the first detonation is uncertain. Previous studies \citep{2011ApJ...734...38W,2012MNRAS.420.3003S} suggest that in a single degenerate scenario, a He shell of 0.1 - 0.2 $\,\mathrm{M_{\odot}}$ is required to trigger the first detonation. But such a massive He shell would lead to the production of titanium, chromium, and nickel, leading to deviation from normal supernova Ia spectra \citep{2010ApJ...719.1067K,2012MNRAS.420.3003S,2021ApJ...922...68S,2022MNRAS.517.5289C}. Here, we present the He shell masses for the systems that undergo double detonation supernovae.

Fig.~\ref{fig:Explosion_Heshell} shows the He masses in systems that undergo double detonation supernovae and He novae. Panel (a) shows the mass of the He shell of the WD at the time of ignition versus the mass of the WD for systems that undergo double detonation supernovae and He novae. In systems undergoing a double detonation supernova, the more massive the accretor, the less helium mass is required for helium ignition. For systems that undergo double detonation supernova, the He shell masses during the first ignition range from $0.05\,\mathrm{M_{\odot}}$ to $0.33\,\mathrm{M_{\odot}}$. Panel (b) of Fig.~\ref{fig:Explosion_Heshell} shows the mass coordinate of ignition ($\mathrm{m_{He}}$) versus the mass coordinate of the C/O - He boundary ($\mathrm{m_{CO/He boundary}}$). We define the C/O - He boundary  as the layer where the He mass fraction falls below 10$\,\%$. The dashed line shows the points where the mass coordinate of the C/O - He boundary equals the mass coordinate of He ignition. We observe the location of the He ignition in most cases is above the C/O - He boundary. 

Among the systems that result in double detonation supernovae, there is a variation in He shell masses. The system for which we find the least massive He shell originates from a binary with a $0.35\,\mathrm{M_{\odot}}$ donor and a $1.1\,\mathrm{M_{\odot}}$ accretor with an initial orbital period of $\sim$ 1.3\,hours begins mass transfer while the donor is still burning He in its core. Before undergoing He ignition, the accretor accumulates $\sim 0.05\,\mathrm{M_{\odot}}$ of He on its surface through steady accretion ($10^{-8} \,\mathrm{M_{\odot}\,yr^{-1}}$). However, the ignition point is not at the base of the He layer. The mass of the envelope above the ignition point is $0.02\,\mathrm{M_{\odot}}$, located well above the base of the He layer but the densities at ignition are larger than the critical density. Hence we classify this system as a double detonation supernova. 

A system resulting in the most massive He shell originates from a binary with a $0.75\,\mathrm{M_{\odot}}$ donor and a $0.7\,\mathrm{M_{\odot}}$ accretor with an initial orbital period of $\sim 1$\,hour, which starts mass transfer while the donor is burning He in the core. The accretor accumulates $\sim 0.33\,\mathrm{M_{\odot}}$ of He through steady accretion ($10^{-8}\, \mathrm{M_{\odot}\,yr^{-1}}$).  Here too, the ignition point is above the base of the He layer, with the envelope mass above the ignition point being $0.31\,\mathrm{M_{\odot}}$, but with densities greater than the critical density. Thus, we also classify this system as a double detonation supernova.

\begin{figure*}
    \centering
    \includegraphics[trim = 0 0 0 0 ,width=1\textwidth]{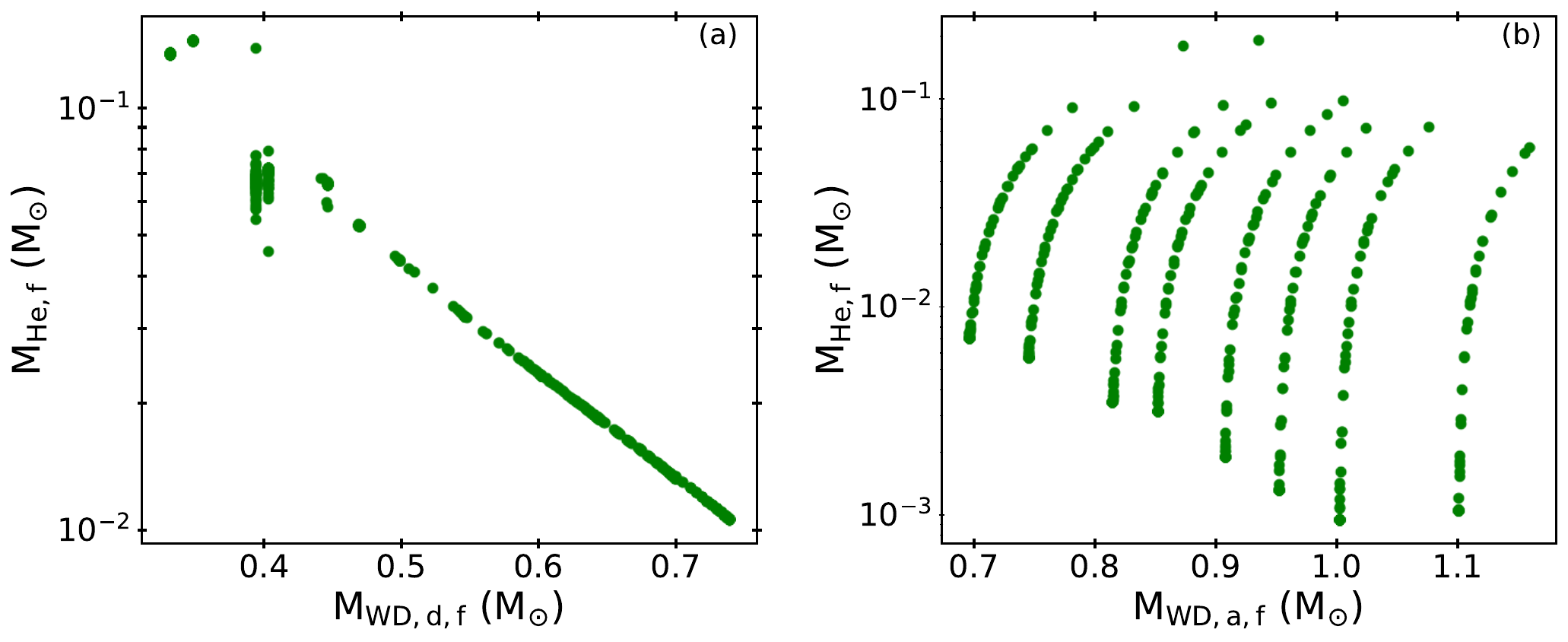}
    \caption{
He shell masses for both the donor WD, which was originally a hot subdwarf (panel a) and the accretor WD, which was originally a WD (panel b) in double white dwarf systems. Panel (a) displays the mass of the donor WD versus the mass of its He shell. Panel (b) shows the accretor WD versus the mass of its He shell. The maximum mass of the accreted He shell is $\sim 0.18\, \mathrm{M_{\odot}}$.
}
    \label{fig:DWDHEshell}
\end{figure*}

\subsection{Runaway velocities} \label{sec:Runaway velocities}

A He-rich star, US 708, was originally observed by \citet{1982ApJS...48...51U} and later observed by SDSS. \citet{2005A&A...444L..61H} measured the velocity of the star and suggested that its high velocity might result from interactions in the Galactic center. \citet{2009A&A...493.1081J} proposed that this velocity could be attributed to the orbital velocity of a hot subdwarf star after its WD companion underwent a Type Ia supernova explosion. This hypothesis was supported by \citet{2015Sci...347.1126G}, and \cite{2015ApJ...804...49B}, who demonstrated that the trajectory of the hot subdwarf star cannot be traced back to the Galactic center. Given the increasing number of hypervelocity runaway stars discovered by Gaia and the upcoming data releases, we anticipate more detections of these stars. 

Since our grid includes models that might lead to WD explosions, we present here the orbital velocities of these donor stars at the time of the explosion of the WD. The orbital velocities range from $\sim 453\,\mathrm{km\,s^{-1}}$ to $\sim 1018\,\mathrm{km\,s^{-1}}$. Fig.~\ref{fig:v_orb} shows the orbital velocity of these donor stars as a function of hot subdwarf mass. The colors represent the current white dwarf mass at the time of the explosion. We note that the mass of the runaway stars may vary depending on the supernova ejecta-donor interaction. \cite{2020A&A...641A..52N} estimated the ejection velocity of US 708 after taking into account the updated proper motion measurements from \citet{2018A&A...616A...1G}. With the inferred ejection velocity to be 894 $\mathrm{km\,s^{-1}}$, and assuming a Chandrasekhar mass supernova, they estimate the mass of US 708 to be in the range $0.34\,\mathrm{M_\odot} < M_{\mathrm{US708}} < 0.37\,\mathrm{M_\odot}$ (see the blue line in Fig.~\ref{fig:v_orb}). In Fig.~\ref{fig:v_orb} the red line and green shaded regions show the  estimated ejection velocities of J2050 (a spectroscopic twin of US 708; \citealt{2017A&A...601A..58Z}), which are \(385 \pm 78 \, \mathrm{km\,s^{-1}}\). From our models, there is one surviving donor with a mass of \(0.63\,\mathrm{M_\odot}\) that lies within their $1\,\sigma$ range, and 69 models within the mass range \(0.5\,\mathrm{M_\odot} < M_{\mathrm{d}} < 0.73\,\mathrm{M_\odot}\) fall within their $2\,\sigma$ range. 

The He star is expected to receive kick velocities of 10-100 $\mathrm{km\,s^{-1}}$ from the ejecta \citep{2023RAA....23h2001L}. However, this is much lower compared than the pre-explosion orbital velocity of the star. Hence, the runway velocities are predominantly determined by the pre-explosion orbital velocities \citep{2023RAA....23h2001L,2024OJAp....7E...7B}. A proto-WD can also be ejected from these binaries if the mass transfer phase is prolonged or begins at a later phase, allowing the donor sufficient time to advance in its evolution toward a WD phase. In Fig.~\ref{fig:v_orb}, the outliers that have higher velocities than hot subdwarfs of the same mass represent these proto-WDs. Some of the faster proto-white dwarfs could potentially account for hypervelocity stars with ejection velocities $\lesssim 1000\ \mathrm{km/s}$, such as those reported by \cite{2023OJAp....6E..28E}, although confirming this would require modeling the impact of supernova ejecta on these stars. 
\subsection{ He shell masses for double white dwarf} \label{sec:Heshell_DWD}

In this subsection, we present the properties of the systems that end up as double white dwarfs. These are systems where either the hot subdwarf star and the WD evolve without interaction or where episodes of accretion on the WD fail to ignite the He shell.

Figure~\ref{fig:DWDHEshell} presents the final He shell mass for both the donor (panel a) and the accretor (panel b) WDs. If the initial orbital period of the binary is large enough to allow for the isolated evolution of the donor, it evolves into a WD with a thick He shell. Panel (a) of Figure~\ref{fig:DWDHEshell} shows the mass of the donor WD versus the mass of its He shell.

About 24$\,\%$ of WDs of all the double white dwarfs undergo accretion either during the thermally pulsing phase or the shell burning phase of the donor, accumulate a range of He masses on the WD, ranging from $\sim$$9.5 \times 10^{-4} \, \mathrm{M}_{\odot}$ to $\sim$$0.18 \, \mathrm{M}_{\odot}$. Panel (b) of Figure~\ref{fig:DWDHEshell} shows the mass of the accretor WD versus the mass of its He shell.

The system that accumulates the maximum He mass ($\sim$$0.18 \, \mathrm{M}_{\odot}$) without He ignition starts with a $0.7 \, \mathrm{M}_{\odot}$ donor, a $0.7 \, \mathrm{M}_{\odot}$ accretor, and an initial orbital period of $\sim$1.3 hours. This system undergoes two phases of mass transfer: the core He burning phase and the He shell burning phase of the donor. This system is explained in detail in Section~\ref{subsec:DWD with massive He shell}. Since the He shell masses are estimated through detailed modeling of binary evolution, these systems provide realistic input for 3D merger simulations, including those that lead to Type Ia supernovae.

\section{Discussion} \label{sec:Discussion}

We simulated the binary evolution of hot subdwarf and WD binaries to study their potential as progenitors of double detonation supernovae, He novae, and double white dwarfs. In this section, we discuss the possible characteristics of double detonation supernovae resulting from our hot subdwarf + WD models, compare our results with previous works, and discuss caveats and limitations inherent in our approach. 

\subsection{Charateristics of double detonation supernovae}

We estimated the He shell masses at ignition for systems that undergo double detonation supernova. From Sect.~\ref{sec:Ignition regions across initial parameter space}, we observe that the He masses required for double detonations range from $\sim0.05\,\mathrm{M_{\odot}}$ to $\sim0.33\,\mathrm{M_{\odot}}$. Previous literature suggests that the spectra from these events will be enriched with titanium (Ti), chromium (Cr), and nickel (Ni), and will not align with the typical SNe Ia spectra \citep{2010ApJ...719.1067K,2010A&A...514A..53F,2020ApJ...904...56G}. According to \citet{2019ApJ...878L..38T}, and \citet{2021ApJ...919..126B}, thin He shells with masses around $0.01\,\mathrm{M_{\odot}}$ are required to replicate SNe Ia spectra. Therefore, these systems are unlikely to represent normal Type Ia supernovae but could result in other peculiar Type I supernova transients. \citet{2022ApJ...925L..12K} estimated the rate of double detonation supernovae with thick He shells to be $4 \times 10^{-6}\,\text{yr}^{-1}$, which is $1\%$ of the complete SNe Ia rate estimated by \citet{2006ApJ...648..868S}.

\subsection{Caveats and limitations}

  In our binary models, we assume conservative mass transfer with no mass lost from the system. This is an initial approximation, as the He retention efficiency of the accretor is still poorly constrained. Further studies are needed to understand the effects of mass loss and angular momentum loss due to mass ejection from the system. Additionally, we assume that any angular momentum gained by the accretor during the accretion process is transferred back to the orbit.  These systems, when in contact, are extremely compact and are commonly expected to be tidally synchronized. Studies such as \citet{2014MNRAS.444.3488F} indicated that tides may be efficient enough to prevent the accretor from spinning up beyond the orbital frequency, and most of the angular momentum gained by the accretor could therefore be transferred back to the orbit. However, other studies have investigated the consequences if the WD can be significantly spun up. For example, \citet{2004A&A...425..217Y} studied rotational spin-up due to He accretion in WDs and found that He shell burning is more stable in rotating WDs compared to non-rotating cases. Similarly, \citet{2017A&A...602A..55N} investigated magnetic and rotating WDs and found that lower mass WDs ($0.54-0.8\,\mathrm{M_{\odot}}$) could accumulate 50\% more mass compared to non-rotating cases before the He ignition. 

We tested the effect of the stellar winds by implementing the wind prescription from \cite{2016A&A...593A.101K}. In our models, stellar winds from hot subdwarfs do not affect the fate of the binary.

Modeling thermonuclear explosions of white dwarfs is inherently a multidimensional problem. \citep{2024A&A...686A.227P}. Our classification between He novae and double detonation supernovae is based on a critical density criterion. We note that the number of systems classified as supernovae changes by less than a factor of two when the critical density threshold is changed from $3.16 \times 10^{5}\ \mathrm{g\,cm^{-3}}$ to $3.16 \times 10^{6}\ \mathrm{g\,cm^{-3}}$.In addition, a He nova is not a final state of the system. Any system classified as He nova can further evolve to either explode the WD or become a double white dwarf. Simulating through He novae involves mass loss assumptions and is computationally expensive. Hence, more work is needed to understand this classification better.  We tested the convergence of our models for the number and grid pattern of double detonation supernovae, He novae and double white dwarfs by incrementally increasing both the time step and mesh resolution.

\subsection{Current and future observations}

The observed system PTF J2238+743015.1 is marked in panel (b) of Fig.~\ref{fig:grid_analysis}. A recent study modeling PTF J2238+743015.1 by \citet{2024A&A...689A.287P} considers the rotational spin-up of the WD due to accretion and finds that the shear heating from this process could raise the surface temperature of the accretor, making it less dense and preventing it from undergoing a double detonation supernova. Consequently, the accretor would fail to detonate the He layers during He accretion. This contrasts with the modeling of PTF J2238+743015.1 by \citet{2022ApJ...925L..12K}, which, using a non-rotating accretor similar to our work, proposes that the binary would end up detonating the WD. While our models do not include the spin-up of the accreting WD or possible associated shear heating, we consider it likely that tides will prevent the WD from spinning up to near critical rotation (see also arguments in \citealt{2021ApJ...922..245B}). Our models therefore allow for the possibility of a supernova detonation for this system, but the relatively low observed mass of the hot subdwarf donor favors a He nova as more likely. 

Another system in panel (b) of Fig.~\ref{fig:grid_analysis} is CD-30°11223 \citep{2012ApJ...759L..25V, Geier2013}. Modeling by \citet{2024MNRAS.527.2072D} proposes that the binary would come into contact before the hot subdwarf star becomes a WD. They also predict that the accretor would undergo detonation in the He layers, which agrees with panel (b) of Fig.~\ref{fig:grid_analysis}.

The observed binary in panel (a) of Fig.~\ref{fig:grid_analysis} is ZTFJ2055+4651, one of the Roche-lobe filling hot subdwarf + WD binaries. Modeling by \citet{2020ApJ...891...45K} suggests that the donor is currently in the Roche-lobe filling stage and will later evolve into a WD, resulting in a double white dwarf system. From our analysis (Fig.~\ref{fig:grid_analysis}), the system could lie in the transition phase between He nova and double detonation supernova. However, since the initial orbital period is longer than the observed orbital period, it is still possible that the system would end up as a double white dwarf, in agreement with \citet{2020ApJ...891...45K}. The other observed systems in panel (g) of Fig.~\ref{fig:grid_analysis} \citep{2000MNRAS.317L..41M,Geier2007,2021NatAs...5.1052P} are at the threshold of multiple possibilities, indicating that the fate of these systems is very sensitive to their initial conditions.

One of the main difficulties in observing these binaries is that the companion is a massive WD. Massive WDs are difficult to observe in electromagnetic observations due to their ultra-compact nature and resulting faintness. However, a sufficiently close massive hot subdwarf is bright ($10-100\,\mathrm{L_{\odot}}$) and can be observed by photometric brightness from Gaia's color-magnitude diagram (for the latest volume completed sample of observed hot subdwarfs, see \citealt{2024A&A...686A..25D}). In addition, due to their short orbital periods, these systems exhibit ellipsoidal modulations due to the tidal deformation of the hot subdwarf. These can be observed using ZTF \citep{2019PASP..131a8002B}, and BlackGEM \citep{2016SPIE.9906E..64B}, OGLE \citep{2015AcA....65....1U}  and Gaia \citep{2016A&A...595A...1G}. Large surveys such as 4MOST \citep{2014SPIE.9147E..0MD}, WEAVE \citep{2012SPIE.8446E..0PD},
and the Milky Way mapper survey included in SDSS-V \citep{2017arXiv171103234K} will provide us with RV variability to detect the binary companion.

\section{Conclusion} \label{sec:Conclusion}

We simulated a grid of binary models containing a hot subdwarf and a WD. Both the hot subdwarf and WD were simultaneously evolved using MESA, incorporating their binary evolution and stable mass transfer via Roche-lobe overflow. Our orbital period spans a wide period range ($36\, \text{minutes} \lesssim P_{\text{orb}} \lesssim 7.2\,\text{hours}$) to accommodate interacting and non-interacting systems. For shorter orbital periods, the binary components come into contact before the hot subdwarf becomes a WD (see Sect.~\ref{Effect of donor on the mass transfer rate} for more details). We identify several phases of mass transfer, including mass transfer during the core He burning phase, the shell He burning phase, and late thermal pulses. The initial parameter space for systems that end up as double detonation supernovae, He novae, and double white dwarfs are shown in Fig.~\ref{fig:grid_analysis}.  In systems where the WD explodes, the runaway velocities of the donor can reach up to $\sim\mathrm{1018\,kms^{-1}}$. Fig.~\ref{fig:v_orb} shows the calculated orbital velocities of donor stars as a function of the hot subdwarf mass. Systems with larger orbital periods tend to evolve into double white dwarfs. We present the most important conclusions here.
\begin{enumerate}
    \item We present the most up-to-date dense grid of MESA models for hot subdwarf + WD binaries that lead to the formation of double detonation supernovae, He novae, or double white dwarfs. Fig.~\ref{fig:grid_analysis} shows the mapping of these outcomes across initial parameter space.
    \item We find that the systems initiate mass transfer during core He burning, shell He burning, and thermally pulsing phase of the donor. In addition, there is also a non-negligible fraction of systems that undergo two phases of mass transfer, which leads to He ignition on the accretor. These systems begin mass transfer during the core He burning phase and later ignite the He shell of the accretor when the mass transfer rate increases during the shell-burning phase of the donor. See Sect.~\ref{Effect of donor on the mass transfer rate} for more details. 
    \item In addition to the runaway velocities of hot subdwarfs, we also estimate the runaway velocities of proto-WDs, which tend to have a higher runaway velocities than that of a hot subdwarf of similar mass. For more details, see Fig.~\ref{fig:v_orb}.
    \item We estimated the He shell masses at the time of He ignition for the systems that undergo double detonation supernova (see Fig.~\ref{fig:Explosion_Heshell}). The minimum mass required for He detonation in our models  is $\sim0.05\,\mathrm{M_{\odot}}$. The white dwarf initially contains only $\sim 2 \times 10^{-3}\,\mathrm{M_{\odot}}$ of He, so the majority of the He must be accreted. In most cases, the ignition point is found to be above the base of the He layer. 
    \item  Double white dwarf resulting from this hot subdwarf + WD binaries have thicker He shells compared to if the WDs had formed from single stars (see Fig.~\ref{fig:DWDHEshell}). This may affect the outcomes of double white dwarf mergers, and so Type Ia supernovae. The WD with the most massive He shell in our double white dwarf systems is a $\sim0.88\,\mathrm{M_{\odot}}$ WD containing a $\sim0.18\,\mathrm{M_{\odot}}$ He shell. For more details, see Sect.~\ref{subsec:DWD with massive He shell}.

\end{enumerate}
We compiled the existing observations of hot subdwarf + WD binaries that fall within the parameter space of our grid and marked them accordingly in Fig.~\ref{fig:grid_analysis}. The marked region aligns well with the detailed modeling of these systems. Therefore, this grid will provide a first-order estimate of the potential outcomes for future hot subdwarf + WD binary observations. Since the properties of our double white dwarf systems are derived from detailed binary evolution calculations, this grid can serve as an input for future 3D merger simulations that aim to model potential Type Ia supernova outcomes.

\begin{acknowledgements}
ASR thanks Stephan Geier, Ken'ichi Nomoto, Ylva Götberg, and Jim Fuller for helpful discussions. We also express our gratitude to the Kavli Foundation for funding the MPA Kavli Summer Program 2023, during which this collaboration became possible. ASR thanks Simon Jeffrey for bringing Ton S 415 to our attention. 

Software: \texttt{MESA} \citep{Paxton2011, Paxton2013, Paxton2015, 2018ApJS..234...34P, Paxton2019, 2023ApJS..265...15J}, \texttt{Matplotlib} \citep{2007CSE.....9...90H}, \texttt{NumPy} \citep{2011CSE....13b..22V}, \texttt{pandas} \citep{mckinney2010}.

{\it Data availability:} All the codes used in this project, including the \texttt{MESA} inlists, are available in the \href{https://zenodo.org/records/13473758}{Zenodo} and \href{https://github.com/AbinayaSwaruba/sdb_WDbinaries.git}{GitHub repository}. The data produced during this study is available upon request.

\end{acknowledgements}

\bibliographystyle{aa} 
\bibliography{paper} 

\begin{appendix} 

\section{Outliers}\label{Appendix:app_ouliers}

\begin{figure}
    \centering
    \includegraphics[width=0.9\columnwidth]{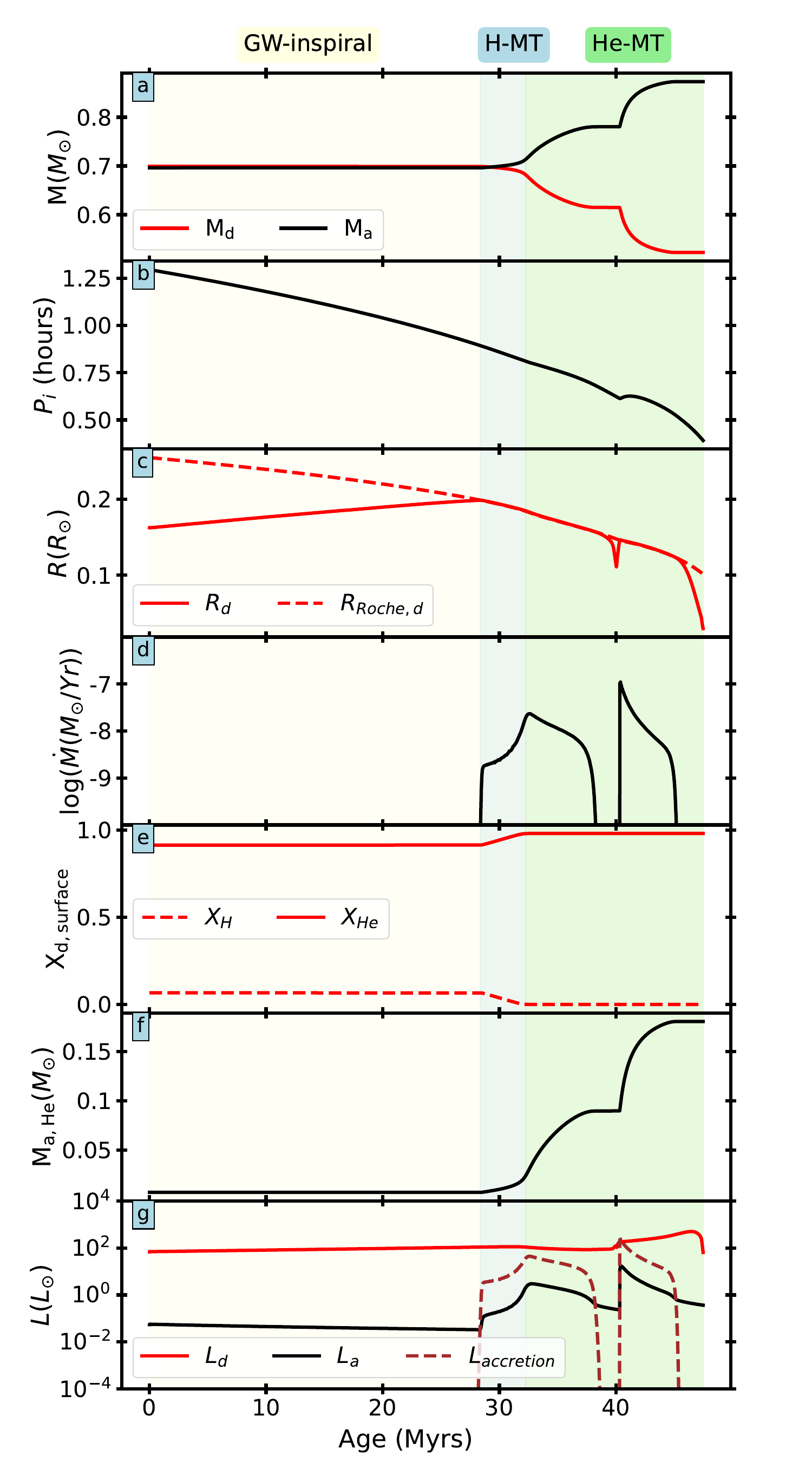}
    \caption{WD undergoes accretion during core He burning and shell burning phases of the donor, eventually accumulating $0.18\,\mathrm{M_{\odot}}$ of He. The panels display the evolution of different parameters as in Fig.~\ref{fig:Explosion}. The background colors yellow, blue, and green represent the gravitational-wave inspiral phase (GW-inspiral), H mass transfer phase (H-MT), and He mass transfer phase (He-MT), respectively. $0.7\,\mathrm{M_{\odot}}$ WD has accreted $\approx 0.18\,\mathrm{M_{\odot}}$ of material producing a $\approx 0.88\,\mathrm{M_{\odot}}$ WD. For more details, see Sect.~\ref{subsec:DWD with massive He shell}. This WD has the most massive He shell in our models.}
    \label{fig:WD with massive He shell}
\end{figure}

\begin{figure}
    \centering
     \includegraphics[width=0.9\columnwidth]{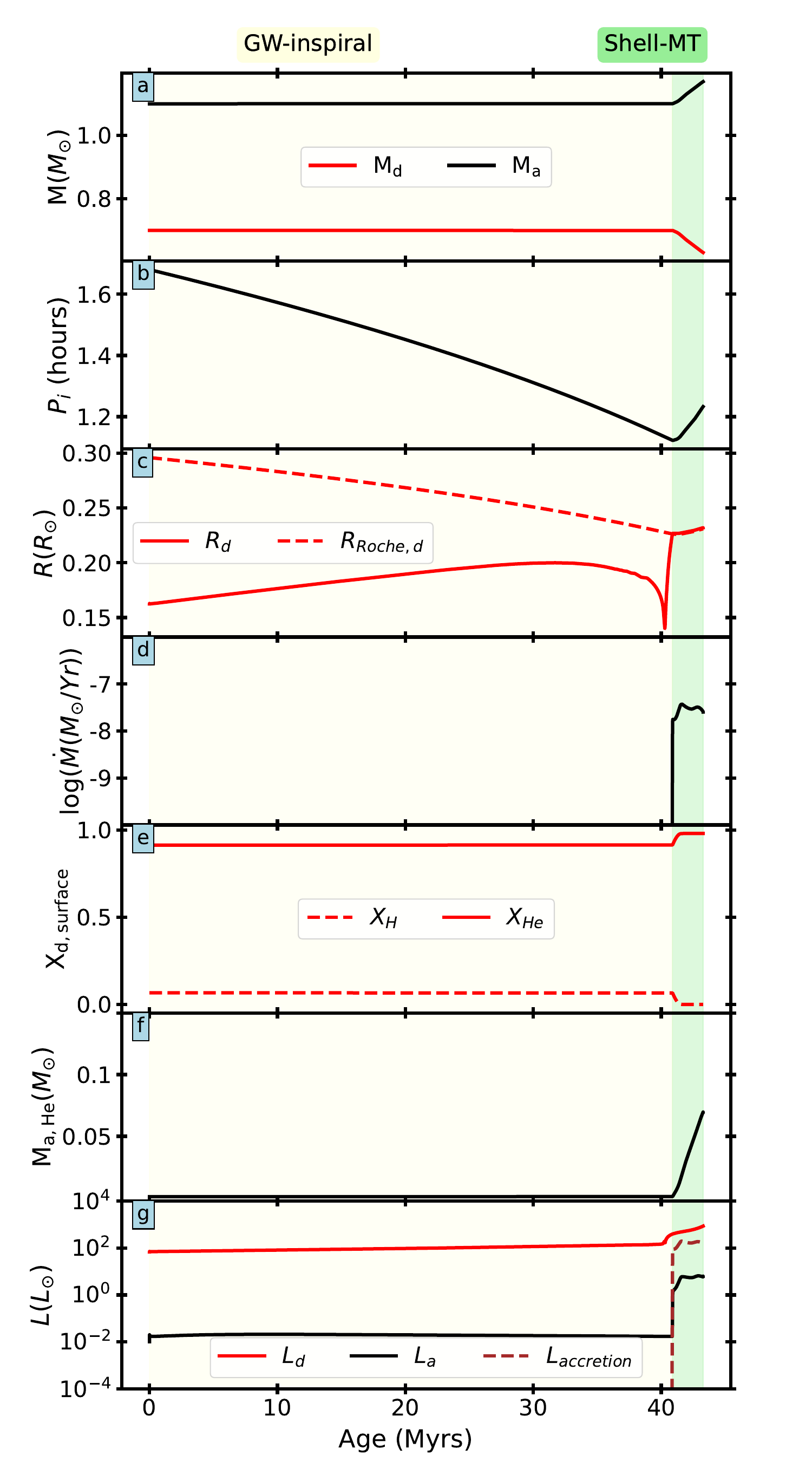}
    \caption{Example of a system where a double detonation supernova occurs during the shell burning phase of the donor. The panels display the evolution of different parameters as in Fig.~\ref{fig:Explosion}. The background colors yellow, and green represent the gravitational-wave inspiral phase (GW-inspiral), and mass transfer phase during the He shell burning phase (Shell-MT), respectively. This system begins interacting during a later expansion phase, where the donor's expansion rate is slower, resulting in lower accretion rates of around $10^{-8}\, \mathrm{M_{\odot}\,yr^{-1}}$. This lower accretion rate leads to He ignition in the dense ($< 10^{6} \, \mathrm{g\,cm^{-3}}$) layers of the WD, ultimately getting classified as a double detonation supernova. For more details, see Sect.~\ref{subsec:wide period supernova}.
}
    \label{fig:Wide period supernova}
\end{figure}

In this section, we present the figure explaining the evolution of properties of the two outlier systems described in section \ref{sec:outliers}. Fig.~\ref{fig:WD with massive He shell} shows the evolution of properties of the binary system, comprising a $0.7\,\mathrm{M}{_\odot}$ donor, a $0.7\,\mathrm{M}{_\odot}$ accretor, and an initial orbital period of $\sim 1.3$ hours. During evolution, the accretor undergoes two episodes of accretion to accumulate $\sim 0.18\,\mathrm{M_{\odot}}$ of He. For more information, see Sect.~\ref{subsec:DWD with massive He shell}. Fig.~\ref{fig:Wide period supernova} shows the evolution of properties of a binary with initial donor mass of $0.7\, \mathrm{M}_{\odot}$, a $1.1\, \mathrm{M}_{\odot}$ accretor, and an initial orbital period of $\sim 1.7\,$hours. This system comes into contact during the later shell-burning phase of the donor and evolves into a double detonation supernova. For more details, see Sect.~\ref{subsec:wide period supernova}.

\end{appendix}
\clearpage
\end{document}